\documentclass[twocolumn]{aastex62}
\usepackage{appendix}

\newcommand{\msun}{M_{\odot}}

\newcommand{\hi}{H{\sc\,i }}
\newcommand{\hii}{H{\sc\,ii }}

\graphicspath{{./}{}}

\received{July 2, 2025}
\revised{September 25, 2025}
\accepted{October 13, 2025}

\shorttitle{Searching for Outflows in NGC 3741}
\shortauthors{Gault et al.}

\begin{document}

\title{Searching for Stellar-Feedback-Driven Outflow Signatures: A Deep Dive into NGC 3741}

\author[0000-0002-2492-7973]{Lexi N. Gault}
\affiliation{Department of Astronomy, Indiana University, 727 East
 Third Street, Bloomington, IN 47405, USA}
 \affiliation{ASTRON, Netherlands Institute for Radio Astronomy, Oude Hoogeveensedijk 4, 7991 PD Dwingeloo, The Netherlands}

\correspondingauthor{Lexi Gault}
\email{lgault@iu.edu}

\author{Liese van Zee}
\affiliation{Department of Astronomy, Indiana University, 727 East
 Third Street, Bloomington, IN 47405, USA}

\author[0000-0002-9798-5111]{Elizabeth A. K. Adams}
\affiliation{ASTRON, Netherlands Institute for Radio Astronomy, Oude Hoogeveensedijk 4, 7991 PD Dwingeloo, The Netherlands}
\affiliation{Kapteyn Astronomical Institute, University of Groningen, Landleven 12, 9747 AD, Groningen, The Netherlands}

\author[0009-0008-5195-7722]{James M. Wells}
\affiliation{Department of Astronomy, Indiana University, 727 East
 Third Street, Bloomington, IN 47405, USA}

\author[0000-0001-5368-3632]{Laura Congreve Hunter}
\affiliation{Department of Physics and Astronomy, Dartmouth College, 17 Fayerweather Hill Road, Hanover, NH 03755, USA}

\author[0000-0001-5538-2614]{Kristen B. W. McQuinn}
\affiliation{Space Telescope Science Institute, 3700 San Martin Drive, Baltimore, MD 21218, USA}
\affiliation{Department of Physics and Astronomy, Rutgers, The State University of New Jersey, 136 Frelinghuysen Road, Piscataway, NJ 08854, USA}
\author{Roger E. Cohen}
\affiliation{Department of Physics and Astronomy, Rutgers, The State University of New Jersey, 136 Frelinghuysen Road, Piscataway, NJ 08854, USA}

\author[0000-0003-4122-7749]{O. Grace Telford}
\affiliation{Department of Physics and Astronomy, Rutgers, The State University of New Jersey, 136 Frelinghuysen Road, Piscataway, NJ 08854, USA}
\affiliation{Department of Astrophysical Sciences, Princeton University, 4 Ivy Lane, Princeton, NJ 08544, USA}
\affiliation{The Observatories of the Carnegie Institution for Science, 813 Santa Barbara Street, Pasadena, CA 91101, USA}






\begin{abstract}

Stellar feedback drives winds and outflows critical to the baryon cycles of low-mass galaxies whose shallow gravitational potential wells make them particularly susceptible to mass and metal loss through outflows. However, spatially resolved observations of stellar-feedback-driven outflows are limited due to their low-surface brightness and transient nature. We present the pilot of a larger multi-wavelength study searching for and quantifying stellar-feedback-driven winds and outflows on both spatially and globally resolved scales for a sample of 40 nearby low-mass galaxies. We search for outflow signatures in the star-forming dwarf galaxy NGC 3741 using new optical imaging and spectroscopy from the WIYN 3.5m telescope in conjunction with VLA 21cm observations and local star formation histories derived from resolved HST photometry. With this extensive dataset, we compare the neutral and ionized gas morphologies and kinematics, calculate mass-loading factors, and investigate spatial variations in the star formation history of NGC 3741. Though the galaxy is experiencing a burst in star formation, we find little evidence of strong outflows and calculate very low mass-loading factors. We suggest that, though star formation activity has increased dramatically in the central region of the galaxy over the last 40 Myr, the star formation rate is not high enough to produce a sufficient amount of high mass stars responsible for fueling outflows. Future analysis of the larger sample will allow us to explore how stellar feedback impacts mass loss on local scales, providing a deeper understanding of the interplay between stellar feedback and the interstellar medium in low-mass galaxies.  

\end{abstract}



\section{Introduction}\label{sec:intro}
Feedback from star formation is a crucial component of the baryon cycle in galaxies --- repeated energy injection from supernovae drives galactic winds that contribute to the mixing of metals within the interstellar medium (ISM) and expulsion of galactic material into the galactic halo (\citealt{Larson1974}, \citealt{Saito1979}). Low-mass galaxies have long been predicted to be particularly susceptible to the effects of stellar-feedback-driven outflows due to their shallow gravitational potential wells (e.g., \citealt{Dekel1986}, \citealt{MacLow1999}, \citealt{Ferrara2000}). The efficiency of stellar feedback and winds in low-mass galaxies is thought to drive fundamental properties such as the low baryon fraction in low-mass systems compared to higher mass galaxies (\citealt{Dave2009}; \citealt{Papastergis2012}). Additionally, these winds carry metals out of low-mass systems, which is thought to drive the mass-metallicity scaling relation (\citealt{Brooks2007}), and can lead to the enrichment of the circumgalactic medium (CGM) and intergalactic medium (IGM) (\citealt{Christensen2018}).  

Simulations and observations of galactic winds and outflows in low-mass galaxies are critical to understanding stellar feedback processes and their impacts on fundamental galaxy properties. Current hydrodynamical simulations of low-mass galaxies have provided constraints for outflow properties including outflow velocities, metal loss and mixing, and mass loss rates. Notably, large scale simulations have predicted high mass-loading factors --- the rate at which mass is carried out of the system relative to the star formation rate (SFR) --- for dwarf galaxies as a result of stellar-feedback-driven outflows (e.g., \citealt{Christensen2016}, \citealt{Nelson2019}, \citealt{Mitchell2020}, \citealt{Pandya2021}). However, these results are in contention with observational studies of nearby low-mass galaxies, which estimate much lower mass-loading factors (e.g., \citealt{Martin1999}, \citealt{2019ApJ...886...74M}, \citealt{Marasco2023}). Focusing on smaller scales, simulations that have tracked the galactic disk and resolved individual supernovae tend to find similarly low mass-loading factors (e.g., \citealt{Fielding2018}, \citealt{Smith2021}, \citealt{Steinwandel2024}). These differences prompt further observational studies on both local and global scales in order to understand the baryon cycle and how mass-loading factors vary with galaxy properties.

Observations of stellar-feedback-driven winds and outflows are often difficult due to the low-surface brightness and transient nature of winds. Many studies have explored samples with global galaxy measurements, analyzing outflow impacts and mass loading on galactic scales (e.g., \citealt{Chisholm2017}, \citealt{2019ApJ...886...74M}, \citealt{Marasco2023}). These global studies, while providing valuable constraints to similarly scaled theoretical studies, tend to wash out any underlying small-scale relationships between stellar feedback and other local properties. Some observational studies have investigated individual galaxies in detail, identifying and quantifying small-scale evidence of stellar feedback such as super bubbles and supernova remnants (e.g., \citealt{Sanchez2015}, \citealt{Gerasimov2022}, \citealt{HamelBravo2024}), but without a full sample of galaxies, the broader conclusions of these studies are limited. Recent spatially resolved observations of outflow properties for higher-mass galaxies (M$_*>$ 10$^{10}$ M$_\odot$) have revealed scaling relationships between local specific star formation rate and the presence of outflows (i.e., \citealt{Reichardt2025}), but few similar studies exist for low-mass galaxies. 

Thus, we present the pilot investigation for an upcoming study exploring the impacts of stellar feedback on the ISM in a sample of 40 nearby low-mass galaxies. In this study, we present a multi-wavelength in-depth analysis of NGC 3741 that combines new spectroscopic ionized gas observations from the SparsePak Integral Field Unit (IFU) (\citealt{Bershady2004}) on the WIYN 3.5m telescope, new optical imaging from the One Degree Imager (ODI; \citealt{Harbeck2014}), neutral gas observations from archival VLA observations (VLA-ANGST; \citealt{VLAANGST}), and star formation histories derived from archival HST observations taken with the Advanced Camera for Surveys Instrument (ACS; \citealt{Ford1998}).

\begin{table}[h]
\caption{Galaxy Properties}
\footnotesize
\begin{center}
    \begin{tabular}{l c c } 
    \hline\hline 
    Feature & Value  \\
    \hline
    RA & 11:36:06.18 \\
    Dec & +45:17:01.1 \\
    Distance$^a$ & 3.21 Mpc \\
    M$_B^b$ & -13.2 \\
    Log(M$_*$/M$_{\odot}$)$^c$ & 7.26 $\pm$ 0.10 \\
    Log(M$_{HI}$/M$_{\odot}$)$^d$ & 8.01 $\pm$ 0.07 \\
    Log(O/H)+12$^e$ & 7.68 $\pm$ 0.03 \\
    Log(SFR$_{FUV}$)$^f$ & -2.26 $\pm$ 0.05 log(M$_{\odot}$ yr$^{-1}$)\\
    \hline 
    \end{tabular}
\end{center}
\tablecomments{Superscripts denote references. $a$. \citealt{Jacobs2009} $b.$ \citealt{Cook2014} $c$. based off 3.6 micron fluxes reported in Dale et al. (2009) $d.$ Derived from this work's reprocessing of VLA-ANGST observations. (\citealt{VLAANGST}) $e$. \citealt{Berg2012} $f$. Derived from FUV magnitude and attenuation from \citealt{Lee2011}.}
 \label{tab:galaxyprops}
\end{table}
 
NGC 3741 was chosen for this pilot study because its ionized gas emission is well matched to the SparsePak field of view (70$''$ x 70$''$), and the galaxy is undergoing a relative starburst event in which we can explore stellar feedback on relevant timescales. NGC 3741 (general properties listed in Table \ref{tab:galaxyprops}) is a dwarf irregular (dIrr) galaxy at a distance of 3.21 Mpc (\citealt{Jacobs2009}). Located on the outskirts of the M81 group at a de-projected distance of 1.65 Mpc from M81 (\citealt{Karachentsev2002}), NGC 3741 is thought to be relatively undisturbed by other galaxies (\citealt{Karachen2003,Karachentsev2004}).

This galaxy has been the focus of previous studies due to its extremely large \hi disk that extends to $\sim$8.3 times its Holmberg radius (\citealt{Begum2005}). Analysis of its \hi rotation curve reveals that it is dark matter-dominated (\citealt{Begum2005}; \citealt{Gentile2007}; \citealt{Allaert2017}), and the neutral gas kinematics in the central disk reveal non-circular motions on the order of 5-13 km s$^{-1}$ (\citealt{Gentile2007}) that may be due to the presence of a central \hi bar and spiral arms (\citealt{Gentile2007}; \citealt{Begum2008}; \citealt{VLAANGST}; \citealt{Banerjee2013}). More recently, \cite{Annibali2022} identify two elongated stellar features in the center of NGC 3741 that coincide with the \hi bar and suggest the presence of a weak stellar bar. Given the relatively young ages of the stars in the stellar bar, they suggest that NGC 3741 either underwent a recent merger or is actively accreting gas from the CGM onto its core. This could be correlated with the recent increase in star formation for NGC 3741, which could fuel stellar-feedback driven outflows.

While these studies have analyzed the neutral gas and stellar component in detail, none have explored the ionized gas kinematics in NGC 3741 and how these kinematics may relate to the features seen in the stellar and neutral gas components. In this paper, we present a multi-wavelength analysis of NGC 3741, provide new insights into the evolution of the galaxy, and establish the methods of analysis for exploring the impacts of stellar feedback on the ISM in a larger sample of nearby low-mass galaxies. 
 
This paper is organized as follows. In Section \ref{sec:observations} we discuss the observations and data processing. In Section \ref{sec:galmorph} we present the results of the general galaxy properties and selection of ionized gas regions used in our kinematic analysis. In Section \ref{sec:SparsePakAnalysis} we present the ionized gas kinematics and metal line measurements. In Section \ref{sec:discussion} we discuss the implications of our findings, and in Section \ref{sec:conclusion} we summarize our findings and discuss the future of the larger study.  

\section{Observations and Data Processing}\label{sec:observations}
In this work, we utilize multi-wavelength data to search for signatures of stellar-feedback-driven outflows in NGC 3741. Spectroscopic IFU observations from SparsePak on the WIYN 3.5m telescope provide ionized gas flux and kinematics. Optical broad and H$\alpha$ narrow-band imaging from pODI on the WIYN 3.5m telescope provide the morphological and flux distribution of the stellar population and ionized gas. Archival VLA radio synthesis observations of the neutral hydrogen 21 cm emission line are used to determine the atomic gas kinematics as well as spatial distribution. Archival HST observations of resolved stellar populations were used to derive star formation histories (SFHs). 

\subsection{pODI Observations}\label{subsec:HaObs}
The One Degree Imager (ODI; \citealt{Harbeck2014}) on the WIYN 3.5m telescope was used when it was partially populated (pODI) to obtain narrow-band and broad-band imaging of NGC 3741 on 2013 March 13. A customized 9-point dither pattern was used for the broad-band filters and a customized 8-point dither pattern was used for the H$\alpha$ filter in order to fill in chip gaps. The source was observed in the $g$ and $r$ broad band filters with 100s exposures and in the H$\alpha$ c6009 Mosaic narrow band filter with 720s exposures. The individual images were stacked resulting in combined exposure times of 900s in $g$ and $r$ bands and 5760s in H$\alpha$. In order to later subtract the stellar continuum from the H$\alpha$ image, an on-off sequence was used in which the filters were switched between the H$\alpha$ c6009 (6563$\text{\AA}$ central $\lambda$) and H$\alpha$ c6009+8nm (6650$\text{\AA}$ central $\lambda$) narrow-band filters, each having 720s exposures.

All images were reduced using the Quick Reduce (\citealt{Kotulla2014}) data reduction pipeline through the One Degree Imager Pipeline, Portal, and Archive (ODI-PPA; \citealt{Gopu2014}) science gateway. This pipeline corrects for CCD artifacts (i.e., crosstalk, saturated pixels, overscan signal), applies the bias, dark, and flat-field corrections, corrects for pupil ghosts, and removes cosmic rays. The reduced $r$-band and H$\alpha$ images were then stacked to create the final images which are shown and described in detail in Section \ref{sec:galmorph} below. The photometric solutions of the pODI images were determined using standard star field observations.

\subsection{SparsePak Observations}\label{subsec:sparseobs}
We obtained spatially resolved spectroscopy of NGC 3741 using the SparsePak IFU (\citealt{Bershady2004}) on the WIYN 3.5m telescope. Arranged in a fixed 70$''$ x 70$''$ square, the SparsePak IFU is composed of eighty-two $4.\hspace{-0.1cm}^{\prime \prime}69$ diameter fibers, with the fibers adjacent to each other in the core and separated by 11$''$ in the rest of the field. All observations were taken with the same bench spectrograph setup, including the 316@63.4 grating, X19 blocking filter and observing at order 8 for velocity resolution of 13.9 km s$^{-1}$ pixel$^{-1}$ (0.306 $\text{\AA}$ pixel$^{-1}$) and wavelength range of 6480 to 6890$\text{\AA}$, centered on 6681.184 $\text{\AA}$. Observations were acquired on the night of 22 April 2017 under clear sky conditions. A single SparsePak field positioned to cover the majority of the H$\alpha$ emission was observed, with a three-pointing dither pattern to fill in the gaps between fibers. For each pointing, three exposures of 900s were taken to detect faint diffuse ionized gas. To remove telluric line contamination and cosmic rays, two 900s exposures of a single SparsePak pointing were taken of nearby blank sky. Due to the three-point dither pattern, some positions near the center of the pointing are observed multiple times. The final analysis for the positions with multiple independent measurements are done using the average of all spectra for that position.

The SparsePak data were processed using the standard tasks in the IRAF\footnote{IRAF is distributed by NOIRLab, which is operated by the Association of Universities for Research in Astronomy, Inc., under cooperative agreement with the National Science Foundation.} HYDRA package. After the data were bias-subtracted, dark-corrected, and cosmic-ray cleaned, the DOHYDRA task was used to fit and extract the apertures from the IFU data. The spectra were wavelength calibrated using a solution created from ThAr lamp observations. The blank sky observations were used to sky-subtract the individual images and a customized Python sky subtraction routine was used to remove sky line residuals. After sky-subtraction, the three exposures of each dither position were averaged together to increase the signal-to-noise ratio (S/N). The data were flux calibrated using standard star observations taken the same night.

To further reduce the noise, the reduced spectra were smoothed by a Gaussian with a sigma of 1 pixel. Peak Analysis (PAN; \citealt{PANDimeo}), an IDL software suite, was used to fit a Gaussian to each fiber spectrum. The recessional velocity of each emission line was determined using FXCOR in IRAF. FXCOR determines the luminosity-weighted mean recessional velocity at every position via cross-correlating the calibrated, sky-subtracted spectra with a template emission line spectrum. The measured FWHMs of the H$\alpha$ line were corrected for instrumental broadening of 48.5 km s$^{-1}$, as measured from the equivalently smoothed ThAr spectra assuming a Gaussian profile. The H$\alpha$ line measurements (fluxes, centers, and velocity dispersions) from PAN were visually inspected and fibers with reasonable fits were mapped to their SparsePak fiber location on the sky. The PAN line fluxes and velocity dispersions along with the velocity field determined by FXCOR are presented and discussed in Section \ref{sec:SparsePakAnalysis}.

\subsection{VLA Observations}\label{subsec:HIdata}

Archival VLA \hi observations in B, C, and D configuration (AO215,VLA-ANGST, \citealt{VLAANGST}) were used in this study. In order to have consistent handling across the full sample, the archival data were reprocessed in AIPS\footnote{The Astronomical Image Processing System (AIPS) was developed by the NRAO.}. The inner 75\% of the bandpass (512 MHz) for each observation block was combined to create a ``channel zero" for the data set. The channel zero file was then flagged for radio frequency interference using the AIPS task TVFLAG prior to flux and phase calibration. After applying the calibration solutions, the line data were corrected for Doppler shifts between observing blocks and then continuum subtracted using the task UVLIN, ignoring line channels for continuum measurement. The observing blocks were combined using natural weighting (robust of 5) for the final data cube using the task IMAGR. The final data cube has a velocity resolution of 2.58 km s$^{-1}$, rms = 1.048 mJy beam$^{-1}$, and beam of 9.47$''$ $\times$ 6.66$''$ with a position angle of 79.5$^\circ$. 

To create zeroth, first, and second order moment maps, the data cube was analyzed in the GIPSY package (\citealt{vander1992}) following standard practice, including a primary beam correction for the moment 0 map. Emission was detected in a velocity range of 175.3-280.9 km s$^{-1}$, and these channels were masked at 3$\sigma$ and then combined for each moment map. 

The resulting zeroth, first, and second order moment maps are presented and described further in Section \ref{sec:HIgasmaps}. From the analyzed data cube, we recover a total \hi flux of 42 $\pm$ 4 Jy km s$^{-1}$, which we use to derive an \hi mass of log(M$_{HI}$/M$_\odot$)=8.01$\pm$ 0.07 assuming a distance of 3.21 Mpc (\citealt{Jacobs2009}). From the VLA-ANGST data processing, they recover a total \hi flux of 32.8 Jy km s$^{-1}$, however, they use a robust weighting of 0.5, which results in higher resolution but lower sensitivity than natural weighting. Both are lower than previous Westerbork Synthesis Radio Telescope (WSRT) observations of NGC 3741, which report a total \hi flux of 59.6 Jy km s$^{-1}$ (\citealt{Gentile2007}). Their data processing resulted in a beam size $\sim1.5$ times the size of our beam, and their data was smoothed by twice the resolution, which could recover lower flux emission on the galaxy's outskirts that is not recovered in our data cube.

\subsection{HST Observations}\label{subsec:HSTdata} 
Archival HST observations taken with the Wide Field Channel of the Advanced Camera for Surveys Instrument (ACS; \citealt{Ford1998}) in the F475W and F814W filters in a single ACS pointing (10915, PI: J. Dalcanton). The ACS instrument has a $\sim$200$''$ x 200$''$ field of view with a native pixel scale of 0$''.$05 pixel$^{-1}$. 

The data reduction and optical image processing were performed in an identical manner to that used in STARBIRDS (\citealt{McQuinn2015}), with a detailed description presented in \cite{McQuinn2010}. In summary, photometry was performed on the pipeline processed and corrected images using \texttt{DOLPHOT}, software optimized for the HST instruments (\citealt{dolphot2000}; \citealt{dolphot2016}). The photometry includes only well-resolved point sources with the same quality cuts as STARBIRDS. More than 10$^6$ artificial stars were inserted individually in the science images and recovered using the same cuts applied to the observed stellar catalogs, providing sufficient number statistics to measure photometric completeness and uncertainties over the color-magnitude diagram when fitting SFHs.

\begin{figure}[t!]
    \centering
    \begin{center}
         \includegraphics[width=\columnwidth]{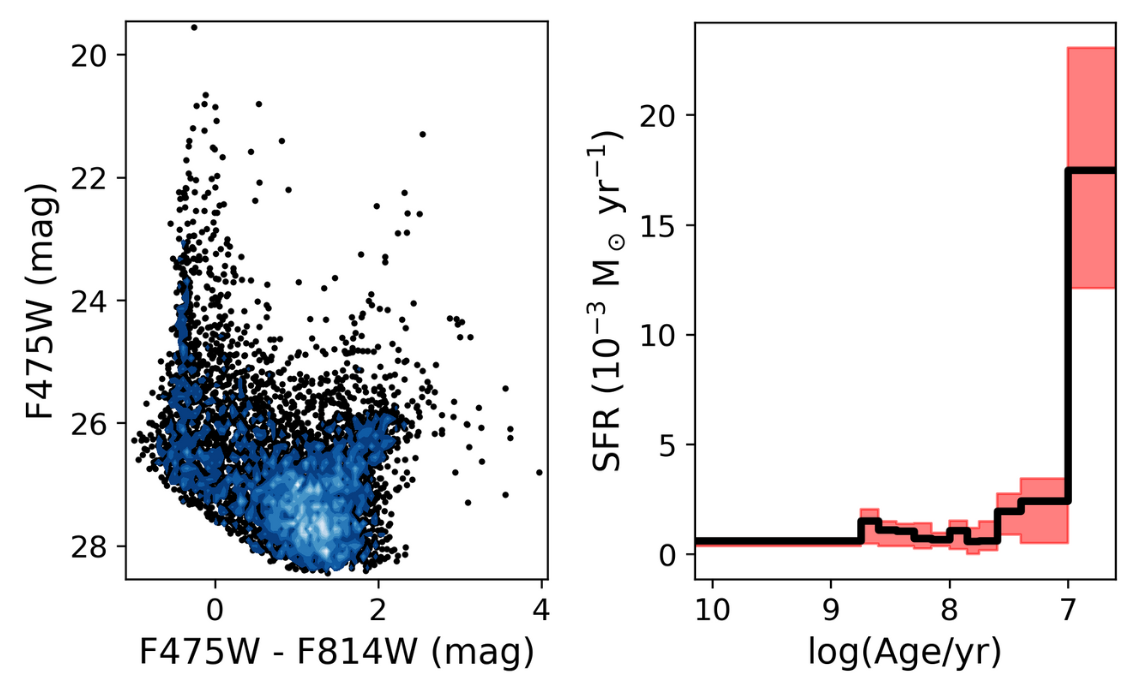}
    \end{center}
    \vspace{-0.4cm}
    \caption{Example CMD (left) for a 400$\times$400 pc box and the corresponding derived SFH (right) with $\leq15$ Myr time resolution in the $t<70$ Myr time bins. See Section \ref{subsec:HSTdata} for specific time binning. Random uncertainties are represented with red shading.}
    \label{fig:CMD}
\end{figure}

The SFHs were derived using color-magnitude diagrams (CMDs) constructed for 400 $\times$ 400 pc$^{2}$ regions across the galaxy. The CMDs were constructed from the ACS observations --- an example CMD is shown in Figure \ref{fig:CMD}. The numerical CMD fitting program \texttt{MATCH} was used to reconstruct SFHs from the resolved stellar populations in each region (\citealt{match2002}). In summary, \texttt{MATCH} uses an assumed initial mass function (IMF) and a stellar evolution library to create a series of synthetic simple stellar populations with various ages and metallicities, and SFH solutions were found for each synthetic CMD. In this case, a Kroupa IMF (\citealt{kroupa2001}) was used and the PARSEC stellar library (\citealt{bressan2012}).The synthetic CMDs were compared to the observed CMDs using a maximum likelihood algorithm to determine if the SFH solution for the synthetic CMD could have produced the observed CMD. The random uncertainties for the SFHs were estimated using a hybrid Markov Chain Monte Carlo simulation (\citealt{Dolphin2013}). The SFHs across the galaxy use time bin edges of log($t$) = [6.6, 7.0, 7.4, 7.6, 7.75, 7.85, 8.0, 8.15, 8.3, 8.45, 8.6, 8.7, 10.15], which correspond to $\sim$15 Myr bins for the first $\sim$70 Myr. For full details see \cite{Hunter2022}.

\section{General Galaxy Morphology}\label{sec:galmorph}
\label{sec:genresults}

\subsection{Stellar Distribution}\label{subsec:stellardist}
\label{sec:starsdist}
\begin{figure}[t!]
\centering
\includegraphics[width=1.0\columnwidth]{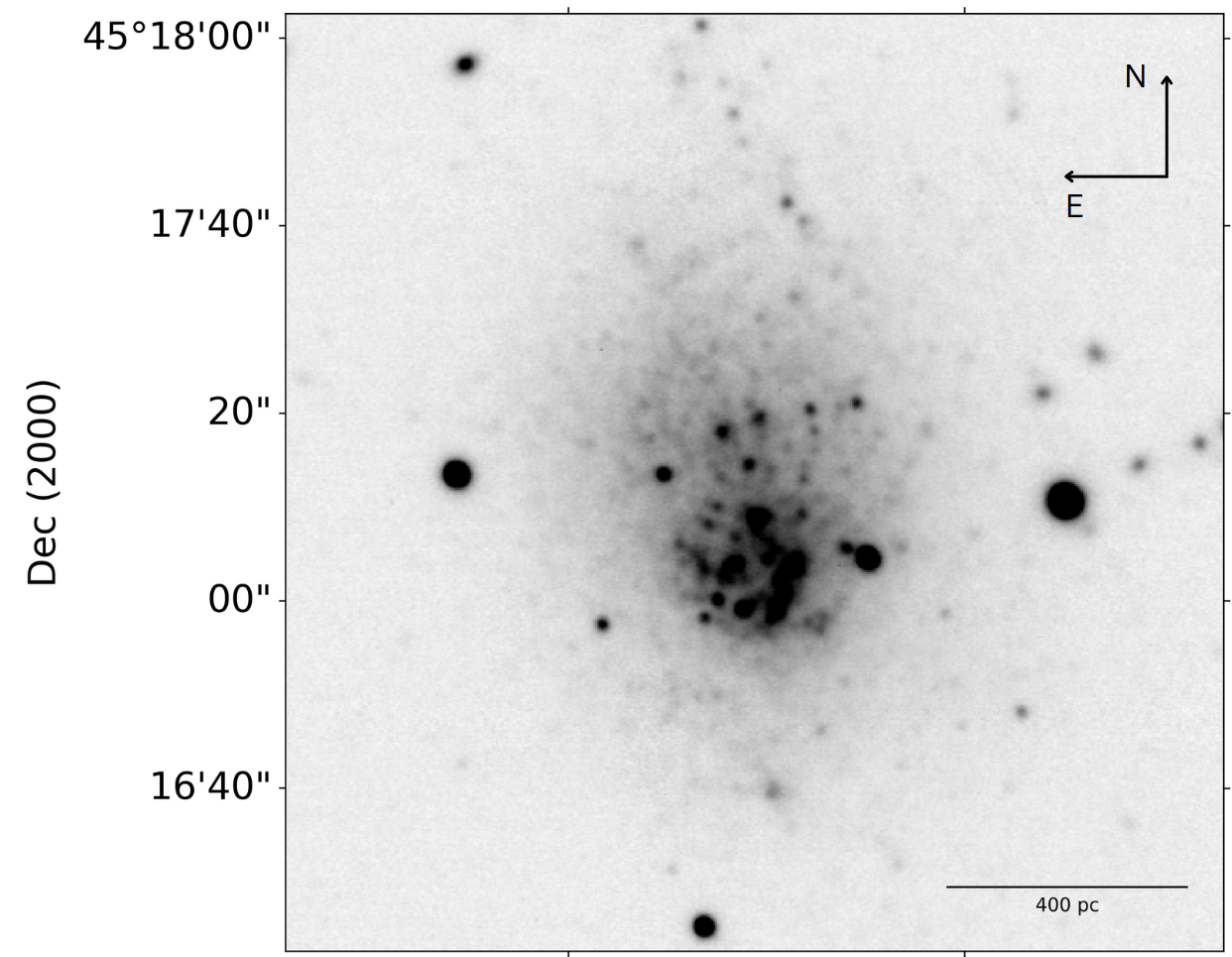}
\includegraphics[width=1.0\columnwidth]{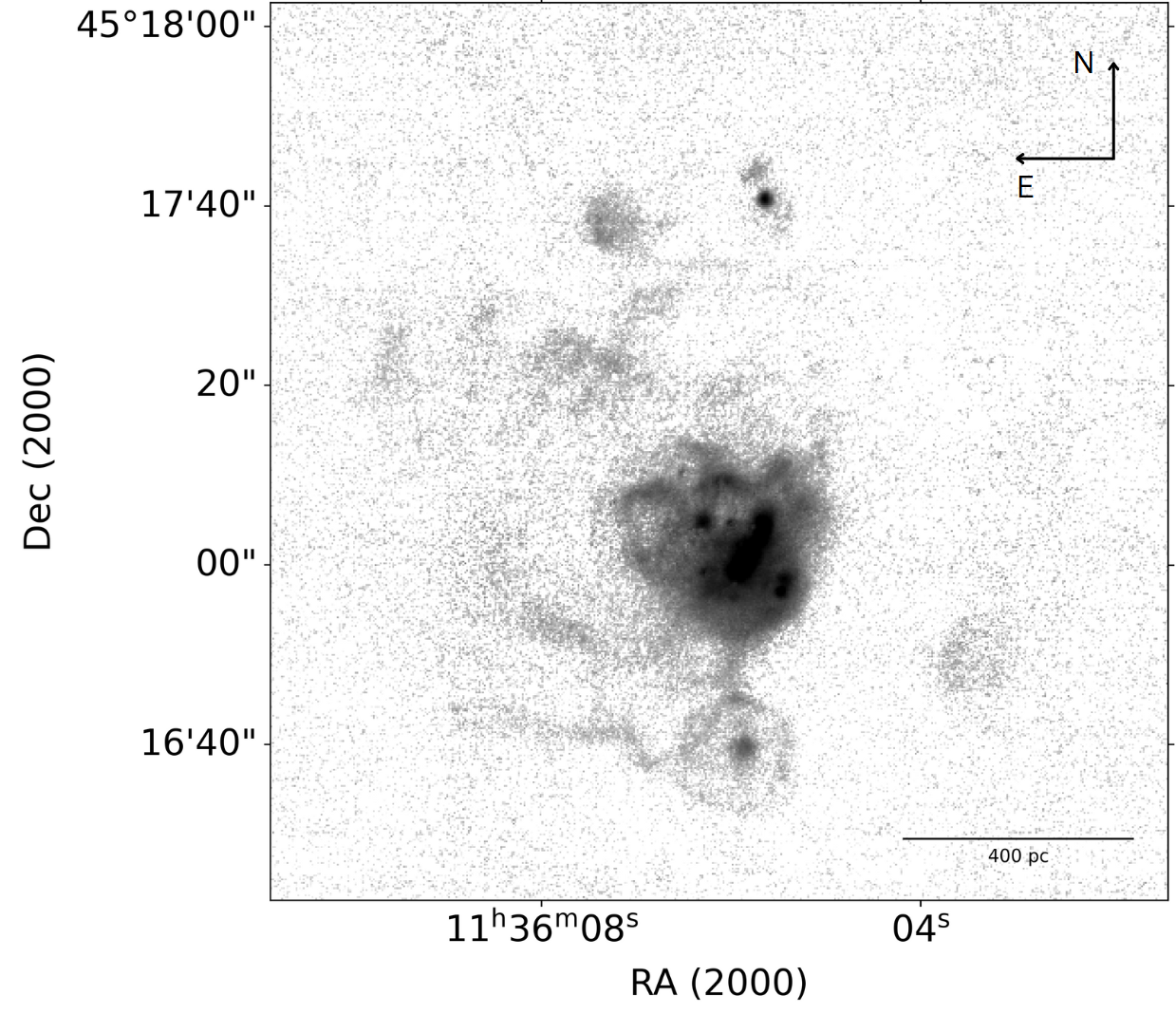}
\caption{ Optical $r$-band (top) and deep H$\alpha$ pODI imaging of NGC 3741. The H$\alpha$ emission is concentrated in a central group of star forming regions with diffuse clouds of emission extending out from the central region. The stellar component shows a bright concentration coincident with the central \hii regions with some faint extensions of stellar material to the north and south.
\label{fig:pODIdata}
}
\end{figure}

The optical $r$-band and the H$\alpha$ pODI images are presented in Figure \ref{fig:pODIdata}. The galaxy's stellar distribution is asymmetric with a higher density of stars in the southern portion of the stellar disk. This has been previously been described as a ``cometary" shape from prior HST observations (\citealt{Karachen2003}). In the Smallest Scale of Hierarchy Survey (SSH; \citealt{Annibali2022}) deep imaging with the Large Binocular Telescope, they identify faint extensions of blue stellar material to the NE and SW, which they suggest may be evidence of a tidally induced bar. The brightest stellar complex corresponds spatially to high-surface brightness H$\alpha$ emission seen in the deep H$\alpha$ image. 

\subsection{H$\alpha$ Optical Imaging}\label{subsec:haoptimaging}
The H$\alpha$ image shows both high-surface brightness and diffuse extended ionized gas structures. The central high-surface brightness H$\alpha$ emission corresponds spatially to the highest surface brightness region of the stellar component, suggesting active star formation in the center of the galaxy. Shown in Figure \ref{fig:pODIdata}, high-surface brightness emission is concentrated in the central region of the galaxy, which coincides with a bright knot of stellar light in the optical image. Outside of the central high-surface brightness regions, there are low-surface brightness features that vary in morphology. Additionally, a superbubble-type feature is evident in the southern part of the galaxy that surrounds a higher surface brightness star forming region.

\subsubsection{\hii Region Selection}\label{subsec:hiiregselect}
In order to identify individual \hii regions and structures of H$\alpha$, we employed a modified version of the Python program \texttt{\texttt{astrodendro}} (\citealt{astrodendro}), adapted to handle our data inputs. The program computes dendrograms, which are hierarchical tree structures, by finding the brightest pixels in the provided dataset and progressively working downward to construct the brightest to faintest detected regions. The \texttt{astrodendro} algorithm takes the following user defined input parameters: minimum flux for a feature, which is a cut to exclude any noise; minimum significance for structures, which sets a minimum difference in flux between one hierarchical structure to the next to ensure the identified regions are separate entities; and a minimum number of pixels that must be included in a structure for it to be considered an independent entity. If a pixel meets the criteria, it merges into an existing structure; otherwise, it initiates a new structure (referred to as a leaf). When adjacent structures connect at a lower-flux pixel, they merge into a branch, forming a nested hierarchy. The process continues until a designated minimum flux value, or “trunk,” is reached, at which point the dendrogram concludes. 

We determined that optimal parameters for the \texttt{astrodendro} program were a minimum flux threshold of 3.58$\times10^{-18}$ erg s$^{-1}$ cm$^{-2}$ per pixel (3$\sigma$ background level), and a minimum of 100 pixels for structure size. Testing lower pixel thresholds introduced significant noise, with diffuse structures and background noise causing the program to generate numerous small nonphysical regions. Conversely, higher pixel thresholds did not yield sufficient detail within the interior regions. These parameters maximized precision while preserving meaningful substructure identification. 

Due to the nature of the dendrogram construction, the low-surface brightness (LSB) regions outside of the central high-surface brightness (HSB) regions were placed into a single dendrogram branch because they do not reach significant local maxima given our input parameters. Thus, we only use \texttt{astrodendro} to identify the HSB \hii regions in the central portion of the galaxy as well as the two additional bright knots in the North. Previous studies that employ \texttt{astrodendro} to identify regions of H$\alpha$ emission tend to use the script to identify bright \hii regions powered by star formation, and then calculate global diffuse ionized gas (DIG) fractions (e.g., \citealt{dellabruna2021}, \citealt{Tacchella2022}). Though the LSB features fall into the lowest branch, they are are spatially distinct regions as seen in Figure \ref{fig:pODIdata}. Thus, we manually select separate regions for each individual LSB region with boundaries based on the lowest branch generated by \texttt{astrodendro}. This effectively creates a surface brightness cut at 1 $\times 10^{35}$ erg s$^{-1}$ $\text{\AA}^{-1}$ arcsec$^{-2}$, where LSB have surface brightnesses less than this cut and HSB have surface brightnesses higher than this cut.

With the HSB regions identified with \texttt{astrodendro} and the manually selected LSB regions, we find 20 distinct regions of H$\alpha$ emission in the galaxy whose properties are listed in Table \ref{tab:hiiregprops}. The radii for each region were measured as half the major axis, which is defined as the longest linear distance across the region. Using the deep H$\alpha$ image, we calculated the flux of each region and subsequently the luminosity.

\begin{table}[t!]
\caption{\hii Region Properties}
\footnotesize
\begin{center}
    \begin{tabular}{l c c c} 
    \hline\hline 
    Region ID & SB$^a$ &
    a$^b$ &  log([SII]/H$\alpha$)\\
    \hline
    \textbf{LSB} \\ 
    1 & 6.42 & 196 & -0.67 $\pm$ 0.09\\
    2 & 7.50 & 192 & -0.76 $\pm$ 0.14\\
    3 & 4.67 & 105 &  -0.66 $\pm$ 0.11\\
    4 & 9.46 & 129 & -0.64 $\pm$ 0.09 \\
    5 & 7.21 & 78 & -0.67 $\pm$ 0.09\\
    6 & 10.05 & 81 & -0.81 $\pm$ 0.16 \\
    7 & 5.92 & 87 & -0.70 $\pm$ 0.10\\
    8 & 8.80 & 107 & -0.70 $\pm$ 0.13 \\
    9 & 7.98 & 127 & -0.71 $\pm$ 0.11 \\
    20$^*$ & 5.59 & 89 & --- \\
    \textbf{HSB} & & & \\
    10 & 12.35 & 85 & -0.83 $\pm$ 0.13 \\
    11 & 11.52 & 57 & -0.96 $\pm$ 0.15\\
    12 & 18.05 & 43 & -0.67 $\pm$ 0.10\\
    13 & 44.17 & 70 & -0.83 $\pm$ 0.14\\
    14 & 31.81 & 145 & -0.94 $\pm$ 0.16\\
    15 & 38.40 & 113 & -0.78 $\pm$ 0.12\\
    16 & 26.35  & 58 & -0.95 $\pm$ 0.17\\
    17 & 76.28 & 69 & -0.93 $\pm$ 0.17\\
    18 & 181.71 & 105 & -1.12 $\pm$ 0.25\\
    19 & 77.50 & 53 & -0.98 $\pm$ 0.18\\
    \hline 
    \end{tabular}
\end{center}

\tablecomments{$a.$ Surface brightness in units of $\times 10^{34}$ erg s$^{-1}$ $\text{\AA}^{-1}$ arcsec$^{-2}$ $b.$ semi-major axis (a) of the region in parsecs. $^*$Region 20 is not covered by the SparsePak pointing and is thus not included in the analysis in Section \ref{sec:SparsePakAnalysis} and onward.}
 \label{tab:hiiregprops}
\end{table}

\begin{figure}[t!]
\centering
\includegraphics[width=0.95\columnwidth]{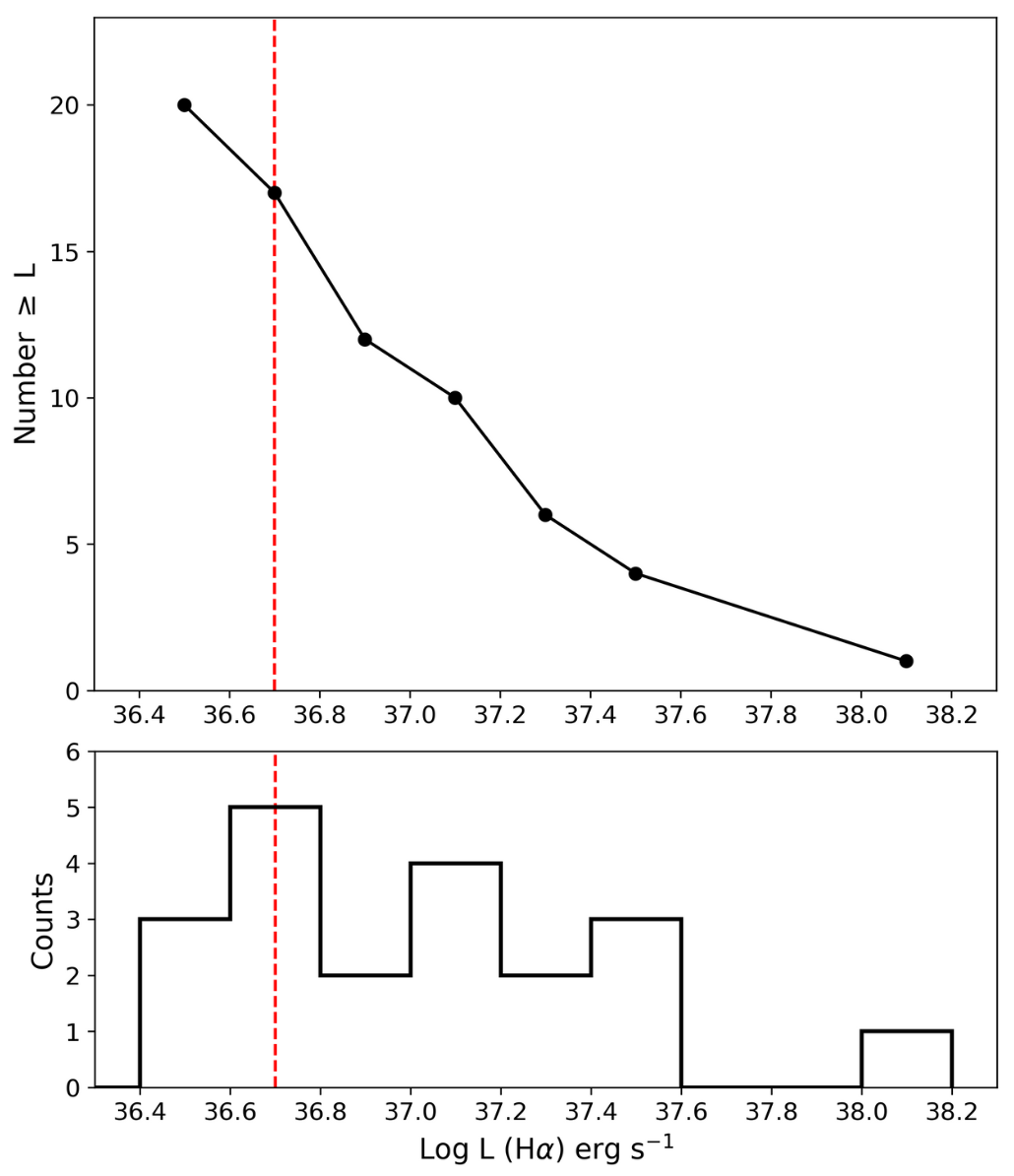}
\vspace{-0.1cm}
\caption{Top panel: Integral \hii region luminosity function derived from the selected regions of H$\alpha$. The vertical dashed line denotes the completeness limit for the galaxy. The luminosity function is steep with a power-law index of -1.39 $\pm$ 0.16. Bottom panel: Histogram showing the distribution of the \hii emission region luminosities from which the luminosity function was derived.
\label{fig:HIIlum}
}
\end{figure}

\subsubsection{\hii Region Luminosity Function}\label{subsubsec:hiireglumfunc}
With the regions identified, we can construct an \hii region luminosity function for the galaxy. Though not all regions can necessarily be defined as star-forming \hii regions as some regions may be shock ionized or ionized by leaky \hii regions, we include all of them for this analysis. We derive an empirical completeness limit by creating a histogram of the fluxes for all 20 identified regions to find a dropoff in counts shown in the bottom panel of Figure \ref{fig:HIIlum}, which resulted in a completeness limit of $\sim4\times10^{-15}$ erg s$^{-1}$ cm$^{-2}$. The derived luminosity function can be seen in Figure \ref{fig:HIIlum}, which was fit to a powerlaw of the form
\begin{equation}
    N(L) = AL^{\alpha}dL,
\end{equation}

where N(L) represents the number of \hii regions with luminosity L, $\alpha$ is the power-law index, and dL is the bin size. Our analysis yielded a luminosity function with a power-law index of $-1.39 \pm 0.16$, consistent with values reported for dwarf irregular galaxies (\citealt{Youngblood1999}; \citealt{elmegreen1999}; \citealt{vanZee2000}). 

\subsection{Neutral Gas Distribution}
\label{sec:HIgasmaps}
\begin{figure*}[ht!]
    \centering
    \begin{center}
         \includegraphics[width=\textwidth]{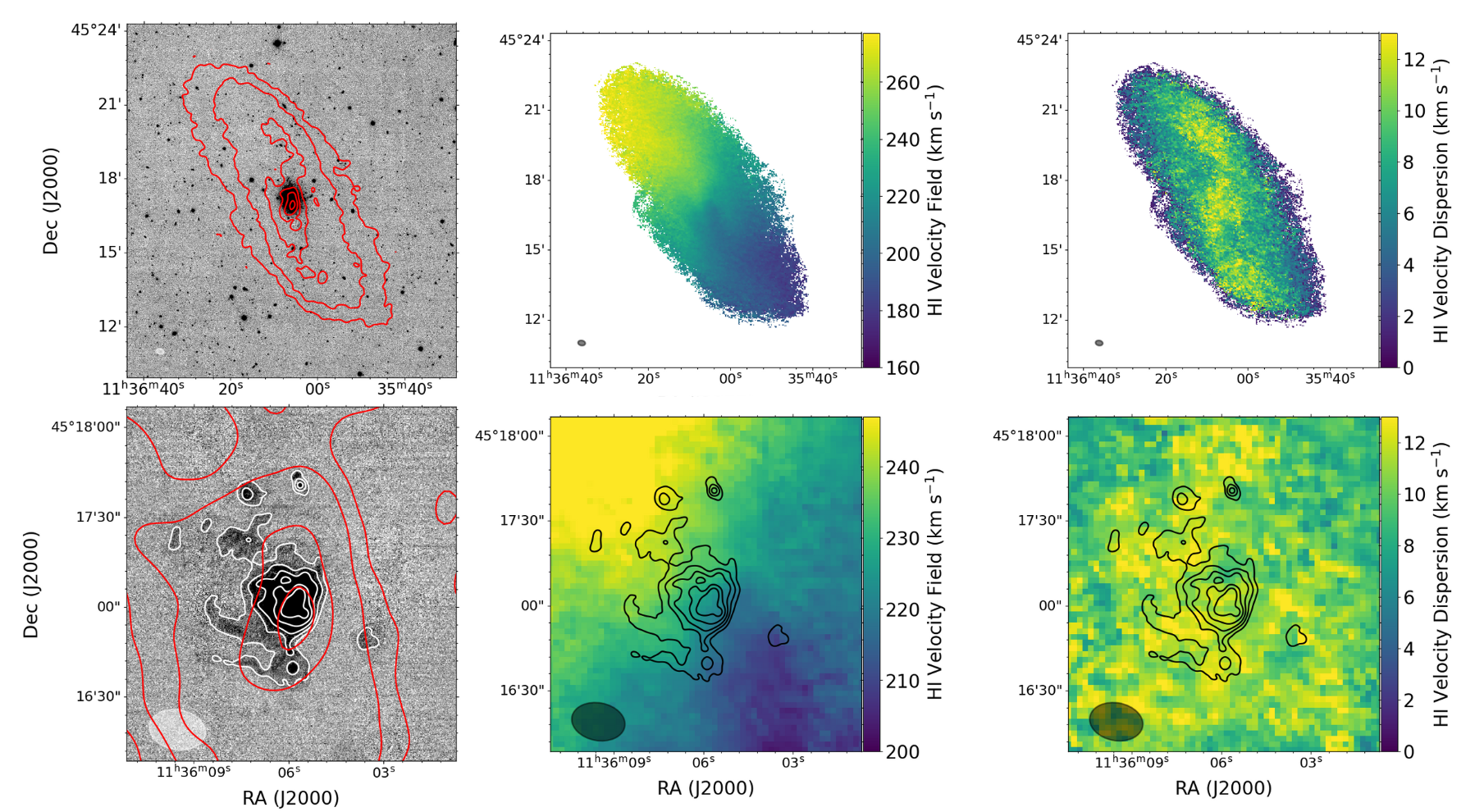}
    \end{center}
    \vspace{-0.4cm}
    \caption{The top panel shows from left to right the \hi column density contours at levels [1.5, 5, 15, 25, 35, and 45] $\times 10^{20}$ cm$^{-2}$ overlaid on the optical pODI $r$-band image, the \hi velocity field of the galaxy, and the \hi velocity dispersion map of the galaxy. The bottom panel shows a zoomed in FOV on the central most region of the galaxy. From left to right this panel shows the H$\alpha$ pODI image with the \hi column density contours overlaid in red and the H$\alpha$ contours overlaid in white and black at levels of [0.2, 0.5, 1, 2, and 4] $\times 10^{-17}$ erg s$^{-1}$ cm$^{-2}$, the \hi velocity field, and the \hi velocity dispersion with the same H$\alpha$ contours overlaid. The ovals in the bottom left of each panel represents the beam size.}
    \label{HImoments}
\end{figure*}

We present the \hi moment maps in the top panel of Figure \ref{HImoments}. The first panel shows \hi column density contours overlaid on the $r$-band pODI image, and the large extent of the \hi disk compared to the stellar component is clear. The second panel shows the \hi velocity field. A clear velocity gradient follows the \hi disk's major axis with evidence of warping in the center of the disk that is also visible the moment 0 map. There is also a warp in the outskirts of the disk, but this is outside of region of interest for this analysis. The third panel shows the \hi velocity widths with regions of broader widths in the center of the galaxy.

In this analysis, we focus on the inner-most portion of the \hi disk that corresponds to the location of the stellar component and the ionized gas emission. This portion of the \hi moment maps are shown in the bottom panel of Figure \ref{HImoments} with H$\alpha$ contours overlaid. The H$\alpha$ distribution aligns with the highest density region of the \hi emission.

\begin{figure*}
    \centering
    \begin{center}
         \includegraphics[width=\textwidth]{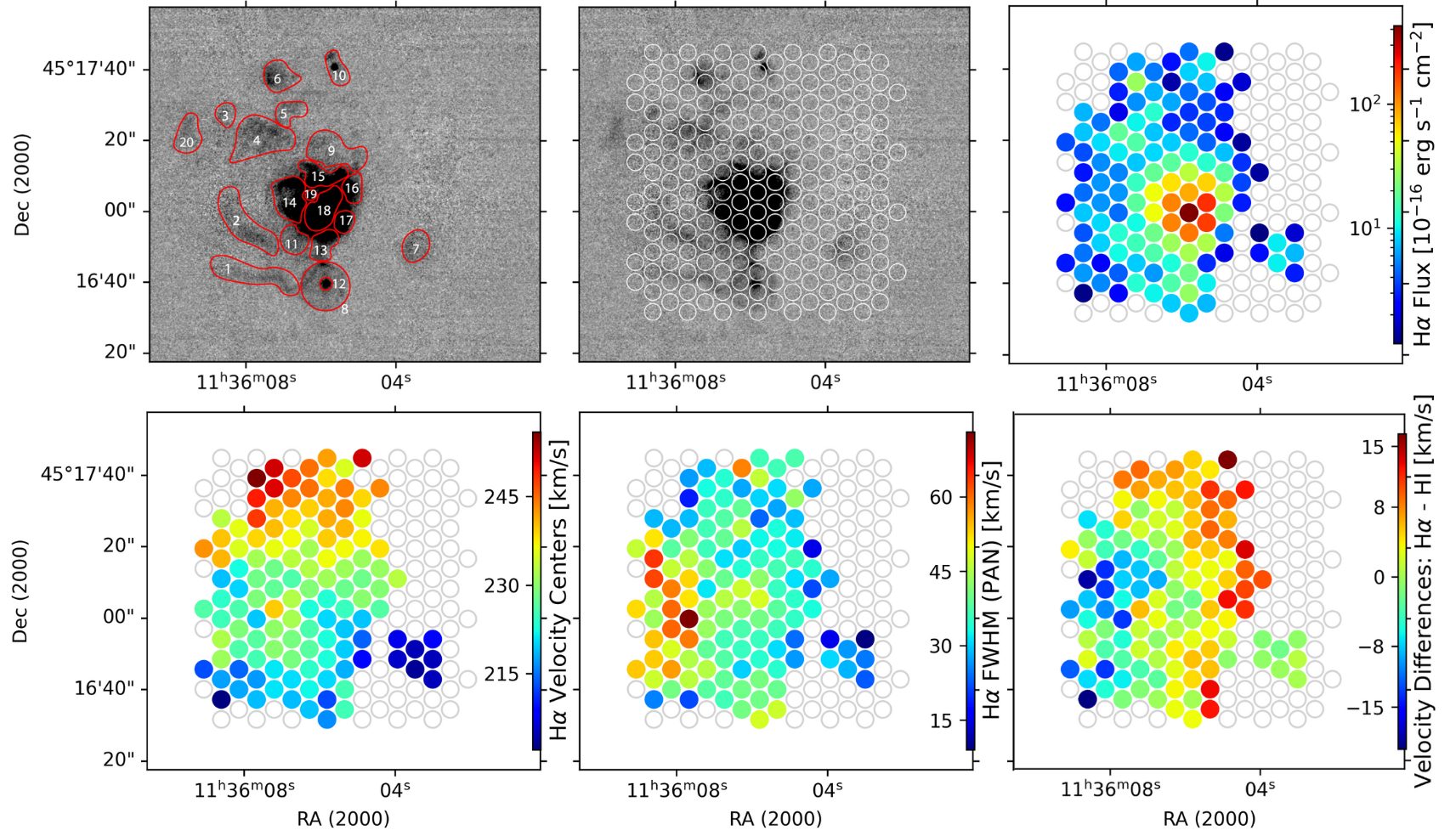}
    \end{center}
    \vspace{-0.4cm}
    \caption{Top row from left to right: optical-continuum subtracted H$\alpha$ image with selected emission regions overlaid, same narrowband image with SparsePak pointing overlaid, and H$\alpha$ flux map. Bottom row from left to right: H$\alpha$ velocity field, H$\alpha$ velocity dispersion, and H$\alpha$ $-$ \hi gas velocity differences (\hi velocity field is shown separately in Figure \ref{HImoments}). The filled circles correspond to the SparsePak pointing fiber locations from where we recover flux.}
    \label{fig:ngc3741}
\end{figure*}

\section{SparsePak Spectral Analysis}\label{sec:SparsePakAnalysis}
\subsection{H$\alpha$ Maps}\label{subsec:Hamaps}
The H$\alpha$ moment maps for NGC 3741 are shown in the bottom panel of Figure \ref{fig:ngc3741}. Of the total 246 SparsePak fibers observed, we recover H$\alpha$ flux in 145 fibers whose H$\alpha$ signal is at least 3$\sigma$ above the spectral noise. Using the SparsePak fibers to perform aperture photometry on the optical image, we find that the H$\alpha$ flux recovered from each spectrum matches the flux recovered in the same spatial regions of the deep H$\alpha$ pODI image. 

The H$\alpha$ velocity field shows a clear velocity gradient, indicative of the underlying bulk rotation of the galaxy with rotation along the galaxy's major axis. The H$\alpha$ FWHMs range from 32 km s$^{-1}$ at the 25th percentile to 44 km s$^{-1}$ at the 75th percentile, with a median FWHM of 38 km s$^{-1}$, which are typical values for low-mass galaxies (\citealt{Marlowe1995}; \citealt{Martin1998}; \citealt{Schwartz2004}). The left side of the galaxy shows a region with large velocity widths, reaching values as high as 73 km s$^{-1}$. These broader spectra align spatially with some of the diffuse ionized gas visible on the left of the H$\alpha$ image (Fig. \ref{fig:ngc3741}). The \hi gas in the galaxy has an average velocity dispersion of $\sim$7 km s$^{-1}$ and a maximum of 18 km s$^{-1}$, indicating that the neutral medium is cooler and likely less disturbed than the ionized gas.

\subsection{Nebular Diagnostics}\label{subsec:nebdiag}
In addition to the H$\alpha$ emission line, our spectral window covers the [NII] 6548 $\text{\AA}$ and 6584 $\text{\AA}$ as well as the [SII] 6717 $\text{\AA}$ and 6731 $\text{\AA}$ emission lines, whose strengths can be used to approximate the electron temperature and electron density of the ionized gas respectively (\citealt{osterbrock2006}). However, we recover very little signal for either [NII] line in the LSB regions, making the measurements of these lines unreliable. For this reason, we proceed with our analysis focusing solely on the [SII] doublet. 

\begin{figure}[t!]
\centering
\includegraphics[width=1.0\columnwidth]{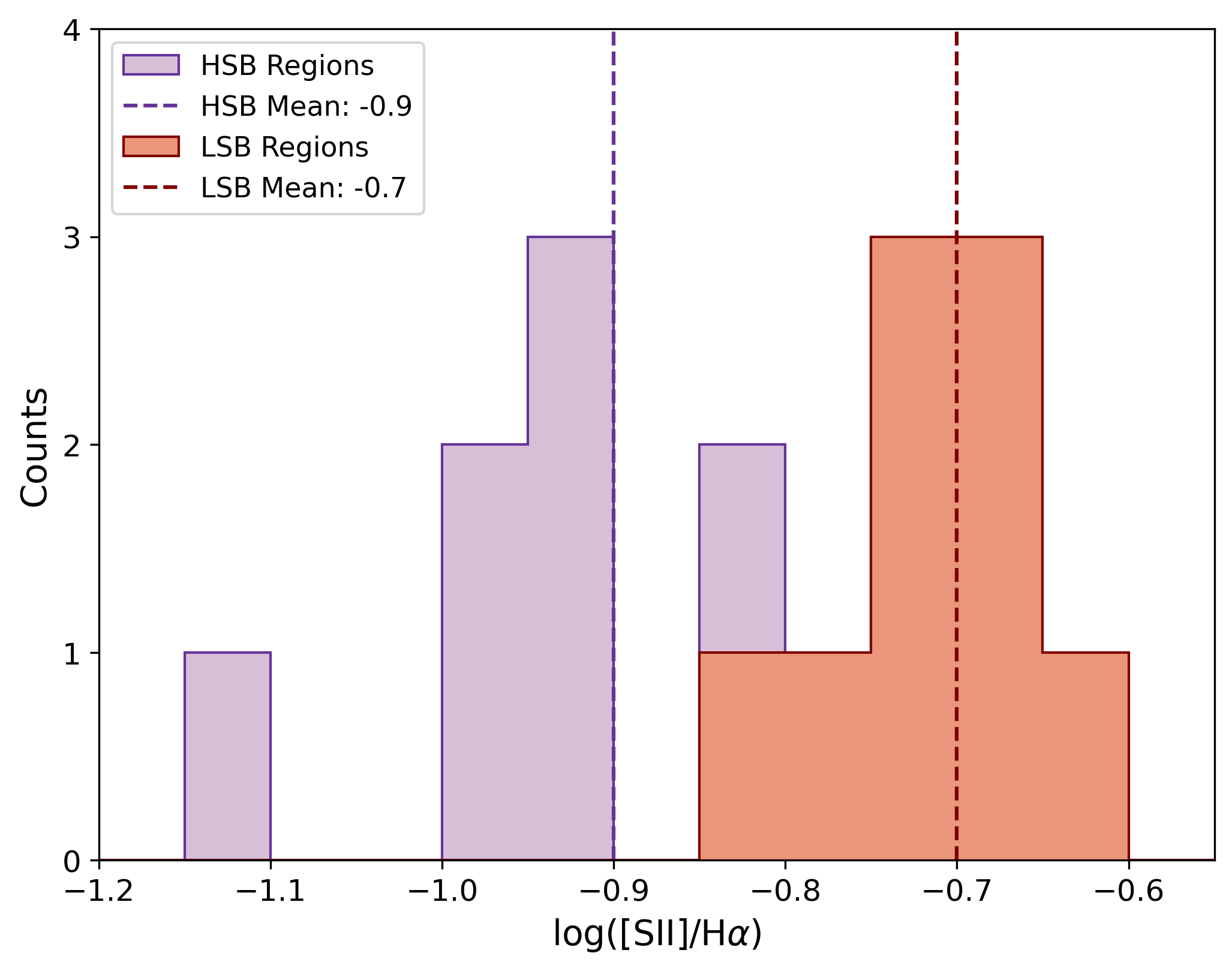}
\includegraphics[width=1.0\columnwidth]{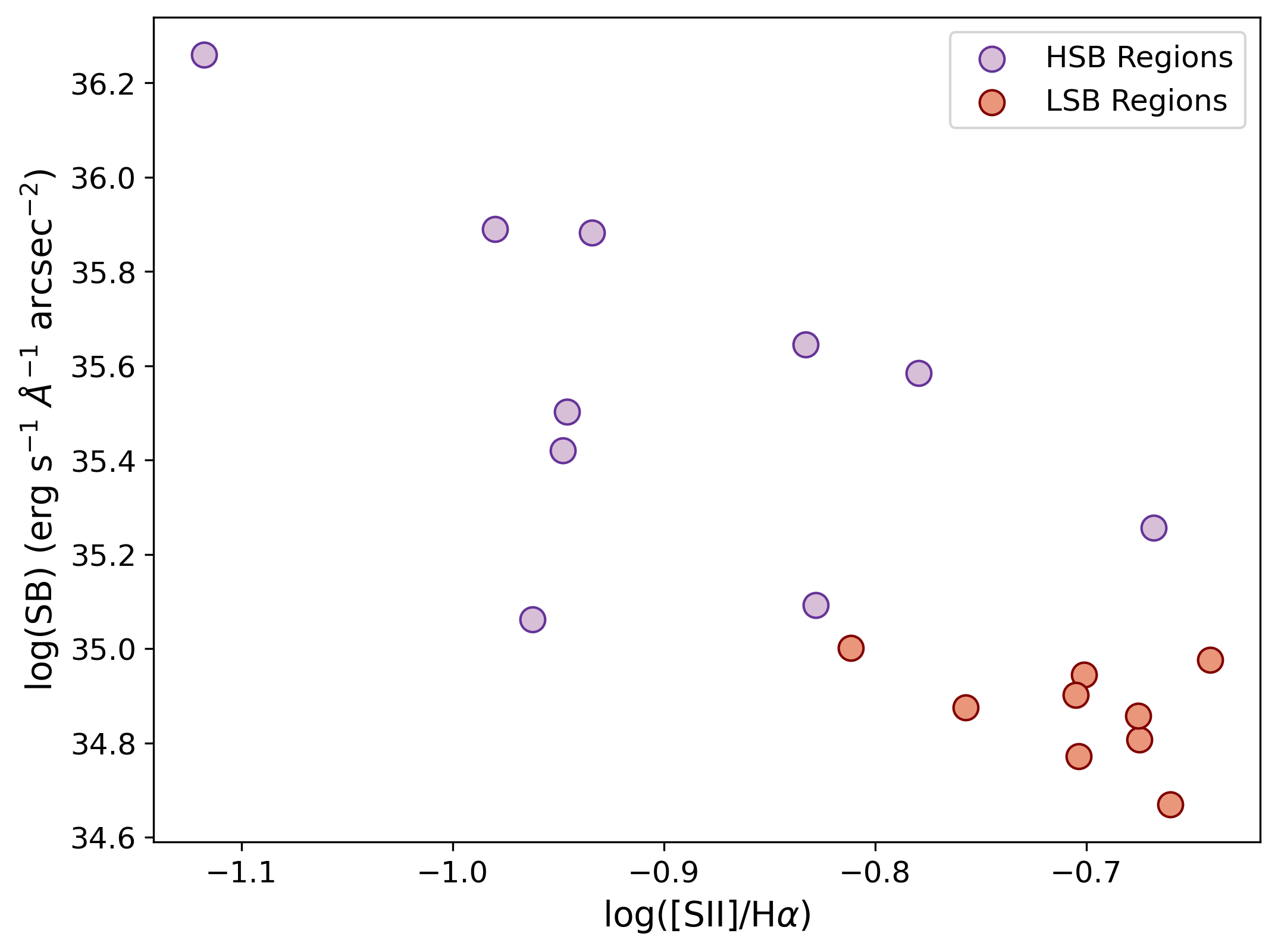}
\vspace{-0.1cm}
\caption{ Top Panel: Histogram of log([SII]/H$\alpha$) for each region of H$\alpha$ emission. The LSB regions are shown in red and the HSB regions are shown in purple. The vertical dashed lines correspond to the average log([SII]/H$\alpha$) for each population: -0.7 for LSB regions and -0.9 for HSB regions. \\ Bottom panel: H$\alpha$ emission surface brightness versus log([SII]/H$\alpha$) plot in which the LSB regions are shown in red and the HSB regions are shown in purple.
\label{fig:SIIHist}
}
\end{figure}

\subsubsection{Metal Line Fitting}\label{subsubsec:metallinefit}
A customized Python routine was used to measure metal emission lines of [SII]$\lambda$ 6716 $\text{\AA}$ and 6731 $\text{\AA}$. For fibers placed on the central complex of H$\alpha$ emission, we recover sufficient signal from the low flux metal lines to fit the reduced spectra directly. However, fibers placed on the diffuse outer regions do not recover enough signal to reliably measure the metal line ratios. Thus, in order to measure the metal emission lines in these regions, we stack the spectra within in each region to increase the S/N. Prior to stacking, the spectra were shifted to the mean centroid of the H$\alpha$ emission line, where the mean centroid is determined as the average value of the FXCOR centroid output for all 145 fibers. The shifted spectra were then summed together for each ionized gas structure. For consistency, we undertook this approach for all regions, resulting in a single measure of the metal lines per region.

Using \texttt{specutils}, an \texttt{astropy} (\citealt{specutils}) Python package built for the analysis of astronomical spectroscopic data, we extract the spectral regions surrounding the emission lines of interest and fit a single Gaussian to each emission line. The flux of each individual line is calculated from the single Gaussian fit, and the individual line fluxes for each [SII] emission line are summed to obtain the total [SII] flux for each region. The measured [SII]/H$\alpha$ flux ratios in low and high-surface brightness regions reported in Table \ref{tab:hiiregprops}

\subsubsection{Metal Line Fitting Results}\label{subsubsec:metlineresults}
From the described fitting, we obtain an average log([SII]/H$\alpha$) of $-0.70$ for the LSB regions and -0.90 for the HSB regions. This amounts to a difference in the mean [SII]/H$\alpha$ ratios of a factor of $\sim$1.6 between the LSB and HSB regions. The top panel of Figure \ref{fig:SIIHist} shows the distribution of log([SII]/H$\alpha$) for both types of regions. We perform a t-test, which is a statistical test to determine if there is a significant difference between the means of two groups, and determine that the high and low-surface brightness regions are statistically different from one another in their [SII]/H$\alpha$ flux ratios. Furthermore, we plot the surface brightness determined from the photometry of each region versus their log([SII]/H$\alpha$) in the bottom panel of Figure \ref{fig:SIIHist} and find that as the surface brightness of a region decreases, its [SII]/H$\alpha$ flux ratio increases. This behavior might be suggestive of different ionization mechanisms powering the H$\alpha$ emission in the high- versus low-surface brightness regions. 

While the central \hii regions are driven by photoionization from the young stellar population, the diffuse ionized gas emission could be shock-ionized. Previous studies suggest line ratios of log([SII]/H$\alpha$) above -0.4 to be shock ionized regions (\citealt{Westmoquette2010}). For NGC 3741, all regions have log([SII]/H$\alpha$) that fall below -0.6. However, when a galaxy has a lower metallicity, the non-photoionization threshold will decrease (\citealt{Westmoquette2010}, \citealt{Dopita2006}) due to inefficient metal cooling keeping the gas warmer and less dense. Therefore, for low metallicity galaxies, like NGC 3741, the threshold for shock ionization may be lower. Thus, the [SII]/H$\alpha$ flux ratios for all regions in the galaxy may be driven down by the galaxy's overall low metallicity of 12+log(O/H) = 7.68 $\pm$ 0.03 (\citealt{Berg2012}). Though we cannot definitively state that the low-surface brightness, diffuse H$\alpha$ emission is shock ionized or ionized by a mix of mechanisms, there is a clear difference in log([SII]/H$\alpha$) values for the central \hii region complex and the diffuse gas regions. 

\subsection{Multi-Gaussian Fitting}\label{subsec:multigausfit}
As discussed, we identify multiple low-surface brightness structures in the H$\alpha$ optical image (Fig. \ref{fig:ngc3741}). Based on the single Gaussian fitting done in PAN, some portions of the galaxy exhibit higher FWHMs compared to the rest of the galaxy as seen in Figure \ref{fig:ngc3741}. These broader spectral widths may be due to underlying wind components, which we explore further with a multiple Gaussian fitting similar to the methods of \cite{Marasco2023}.  

\begin{figure*}
    \centering
    \includegraphics[width=1\linewidth]{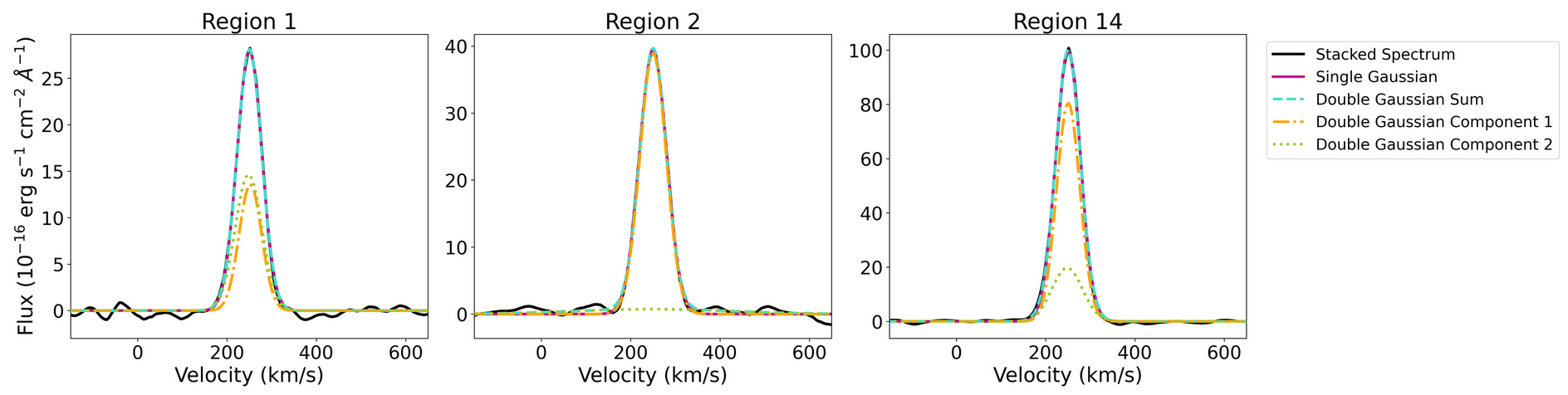}
    \caption{Single and double Gaussian fits plotted over the spectra for regions 1, 2, and 3. The single Gaussian is a statistically preferred fit in each case.}
    \label{fig:gaussianfits}
\end{figure*}

The multiple Gaussian fitting was done using \texttt{lmfit} (\citealt{lmfit}), a Python package built for non-linear optimization and curve fitting. We performed single and double Gaussian fits on the stacked spectrum for each identified region. For the double Gaussian fitting, we constrain the primary component to a velocity width equivalent to the median FWHM of all spectra as measured in Section \ref{subsec:sparseobs}. We assume any underlying wind component will be symmetric about the centroid of the single Gaussian fit, and thus constrain the centroid of both Gaussian components to within 2 km s$^{-1}$ of the centroid found by the single Gaussian fit, consistent with the typical centroid uncertainty. Examples of the single and double Gaussian fits are shown on three H$\alpha$ emission regions in Figure \ref{fig:gaussianfits}.

In comparing the Bayesian information criterion (BIC) between a single Gaussian fit and a double Gaussian fit, all spectra, including both stacked and individual, are better fit with a single Gaussian. This result suggests that within NGC 3741, there are no significant underlying winds within the ionized gas.  

\subsection{H$\alpha$-\hi Gas Kinematics Comparison}\label{subsec:hahicomp}
Another way to uncover underlying bulk motions in the galaxy is through comparing the ionized and neutral gas kinematics. Previous studies have compared ionized and neutral gas kinematics to identify outflows and have found velocity differences between the ionized and neutral gas up to 60 km s$^{-1}$ (\citealt{vanEymeren2009a,vanEymeren2009c,vanEymeren2010}). For NGC 3741, both the H$\alpha$ and \hi velocity fields exhibit fairly ordered bulk rotation. While there is a warp in the \hi disk visible in Figure \ref{HImoments}, this portion of the analysis is focused on the inner portion of the galaxy where \hi disk is undisturbed, with a well-ordered velocity field. To uncover any underlying velocity offsets between the ionized and neutral gas, we subtract the \hi velocity centers from the H$\alpha$ velocity centers in the same regions.

Our spatial resolution for both the SparsePak  and VLA observations are fairly comparable, so we do not perform any smoothing to either map. Using the sky positions of the SparsePak fibers, we extracted neutral gas velocities at each fiber location using the \hi velocity field. The corresponding local \hi velocity was then subtracted from the ionized gas velocity at each fiber location, resulting in the velocity differences shown in the bottom right panel of Figure \ref{fig:ngc3741}. We reveal several regions with bulk motions offset between the neutral and ionized gas. While the velocity offsets are modest, from $-$26 km s$^{-1}$ to 15 km s$^{-1}$, they include both blue-shifted and red-shifted gas. The greatest offsets align with diffuse ionized gas and correspond to regions of higher velocity dispersion in both the ionized and neutral gas, but in the central regions with higher SF activity, there is little difference between the neutral and ionized gas velocities. These velocity differences could indicate wind motions in the diffused ionized gas; however, we cannot directly determine whether these offsets are due to line-of-sight motions of the ionized gas or of the neutral gas. Though, given that the \hi disk appears orderly on the scale of the H$\alpha$ emission, we proceed under the assumption that the velocity offsets are due to non-circular motions of the ionized gas.

\section{Calculating and Interpreting Mass-loading Factors}\label{sec:discussion}
Our analysis has revealed that NGC 3741 has no significant underlying wind components in the H$\alpha$ emission line profiles, but we have recovered modest velocity differences between the ionized gas and neutral gas kinematics. Additionally, we have measured [SII]/H$\alpha$ flux ratios in both the high and low-surface brightness regions, suggesting different ionization mechanisms may be driving the emission in these regions. If we interpret the velocity offsets between the ionized and neutral gas components as expanding wind velocities for the ionized gas, we can estimate mass-loading factors --- the ratio between the mass outflow rate and the star formation rate --- across the disk of the galaxy. In this section, we outline the general method for calculating mass-loading factors, detail our varied measurements, and interpret our results. 

\vspace{1cm}

\subsection{Measuring Mass-loading Factors}
In general, mass-loading factors are defined as $\eta \equiv \dot{M}$/SFR, where $\dot{M}$ is the mass outflow rate and SFR is typically taken as the global SFR for the galaxy. In order to estimate a mass outflow rate, we first must estimate the mass contained within the wind. Previous studies have used the luminosity of the wind component of the H$\alpha$ emission line (e.g., \citealt{cresci2017}; \citealt{marasco2020}; \citealt{Marasco2023}) to estimate the mass assuming that each individual region is a collection of ionized gas clouds with the same electron density, $n_e$, and that the ionization conditions are the same across the entire region. We estimate the total H$\alpha$ luminosity, $L^{H\alpha}_{wind}$, using the integrated H$\alpha$ flux recovered from the stacked spectrum in each region of interest described in the following subsection. Following the method described in \cite{Marasco2023}, we compute the mass using the following:
\begin{equation}
    M_{wind} = 3.2 \times 10^5 \left(\frac{L^{H\alpha}_{wind}}{10^{40} \text{erg s}^{-1}}\right)\left(\frac{100 \text{ cm}^{-3}}{\text{n$_e$}} \text{}\right)\text{M}_\odot
\end{equation}
The electron density for each region was determined from the [SII]$\lambda6716$/[SII]$\lambda6731$ flux ratio, using the relation described in \cite{sanders2016}.

From here, we can estimate the mass outflow rate, $\dot{M}_{wind}$. As an approximation of the wind velocity, $v_{wind}$, we use the average magnitude of the ionized$-$neutral gas velocity offsets in each region. The physical size of the wind, $r_{wind}$, is determined by the physical extent of the region of interest. Again following \cite{Marasco2023}, the resulting mass outflow rate is calculated as 
\begin{equation}
    \dot{M}_{wind} = 1.3 \times 10^{-9} \left(\frac{v_{wind}}{\text{km s}^{-1}}\right)\left(\frac{M_{wind}}{\text{M}_\odot}\right)\left(\frac{\text{kpc}}{r_{wind}}\right) \text{M$_\odot$ yr}
\end{equation}
This value is then divided by the SFR to obtain a dimensionless mass-loading factor. However, with our dataset, there are multiple approaches to measuring SFRs for each region, which could impact the resulting mass-loading factors we measure.

\begin{figure}[h!]
    \centering
    \includegraphics[width=\columnwidth]{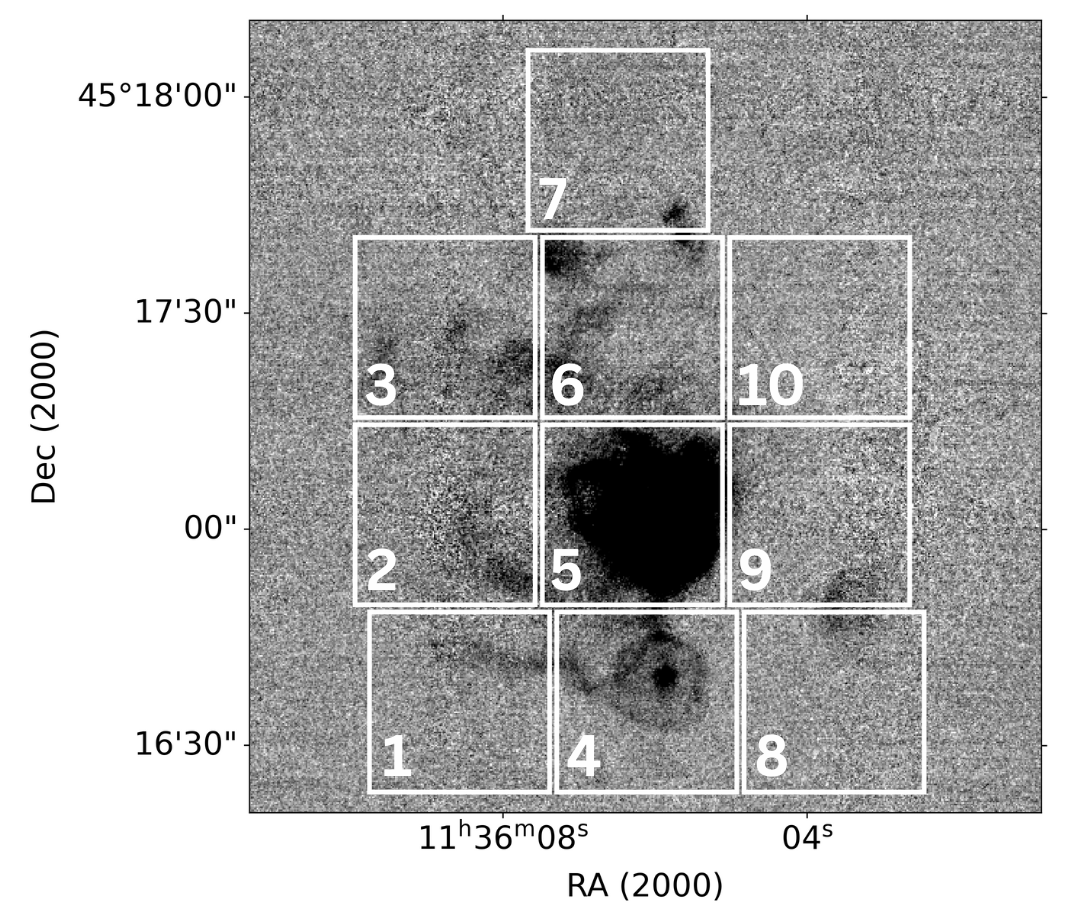}
    \begin{center}
        \includegraphics[width=\columnwidth]{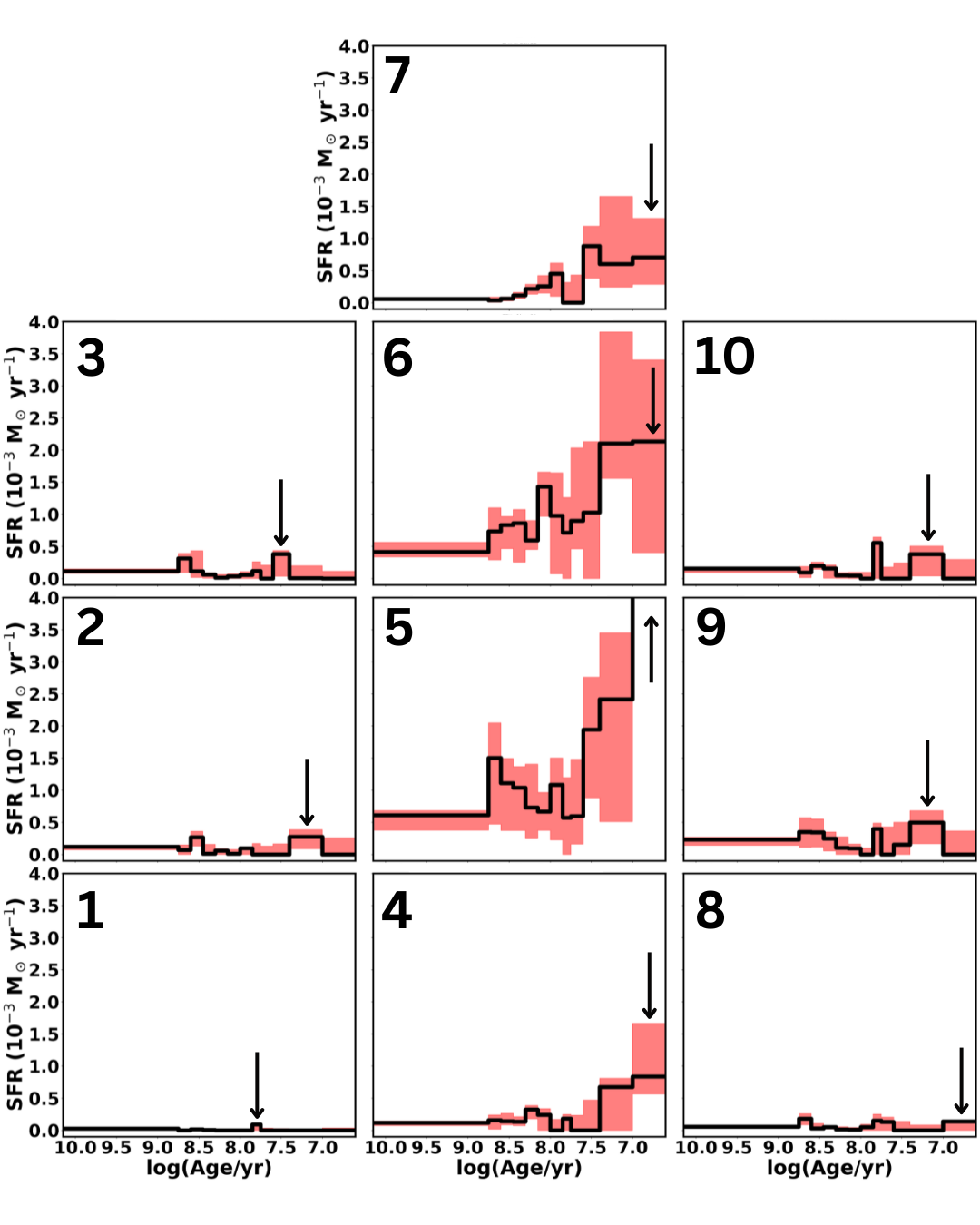}
    \end{center}
    \caption{Top panel: Deep H$\alpha$ pODI image with 400 pc x 400 pc SFH boxes overlaid. Bottom panel: SFHs for each box are plotted. The SFHs have $\leq$15 Myr time resolution in the $t<70$ Myr time bins. See Section \ref{subsec:HSTdata} for specific time binning. The red shading represents the random uncertainties. All SFHs are plotted with the same y-axis scale. Region 5 is shown in Figure \ref{fig:CMD} with full y-axis scale. The black arrows point to the SFRs used to calculate the mass-loading factors for each box, which are listed in Table \ref{tab:massload}.}
    \label{SFHs}
\end{figure}

\subsection{SFR Measurements}
Within our dataset, we can measure the SF activity in local regions across the disk of the galaxy in two ways. We first estimate SFRs for the H$\alpha$ emission regions using the H$\alpha$ luminosity that is estimated from the total H$\alpha$ flux recovered from each region's stacked spectrum. The SFR for each H$\alpha$ emission region is estimated from the region's H$\alpha$ luminosity using the conversion factor from \cite{Kennicutt1998}:
\begin{equation}
    \text{SFR}(M_\odot \text{ yr}^{-1}) = \frac{L(H\alpha)}{1.26 \times 10^{41}\text{ ergs s}^{-1}}
\end{equation}
However, it is important to note that for SFR(H$\alpha$) $< 10^{-4}$ M$_\odot$ yr$^{-1}$, star formation is predominantly stochastic, and H$\alpha$ luminosity is not necessarily a reliable tracer of the star formation activity in a region (\citealt{Weisz2011}). All H$\alpha$ emission regions in the galaxy have SFR(H$\alpha$) below this threshold. 

Because the estimated SFRs in the H$\alpha$ emission regions are very low and thus dominated by stochastic star formation, we want to estimate additional mass-loading factors using local star formation histories that are more indicative of the true star formation activity across the galaxy. The SFHs were derived for \cite{LauraPhDThesis} and Hunter et al. (in prep), in which the galaxy was split into 400 $\times$ 400 pc$^{2}$ regions (26$''$ $\times$ 26$''$ for NGC 3741). This spatial scale was chosen because simulations and theoretical work predict the range over which superbubbles input momentum into the ISM to be one to a few times the scale height of the galaxy (\citealt{Kim2017}; \citealt{gentry2017}), and superbubbles are predicted to be a major driver of turbulence. Typical scale heights for dwarf galaxies are 200-600 pc (\citealt{bacchini2020}), so the selected region size ensures that the measured star formation activity is distinguishable on local scales. Further criteria for region selection can be found in \cite{Hunter2022} and \cite{Hunter2023}.

SFHs for 10, 400 $\times$ 400 pc$^2$ boxes across the H$\alpha$ emission in NGC 3741 along with their corresponding spatial locations on the galaxy are shown in Figure \ref{SFHs}. We estimate mass-loading factors for each SFH box with the SFRs being the most recent star formation event for each box since these events are on timescales that correspond to the current H$\alpha$ emission.

Using these SFR measurements, we can explore mass-loading across the disk of the galaxy on local scales.

\subsection{Mass-loading Measurements in NGC 3741}
\begin{figure}[h!]
\centering
\includegraphics[width=1.0\columnwidth]{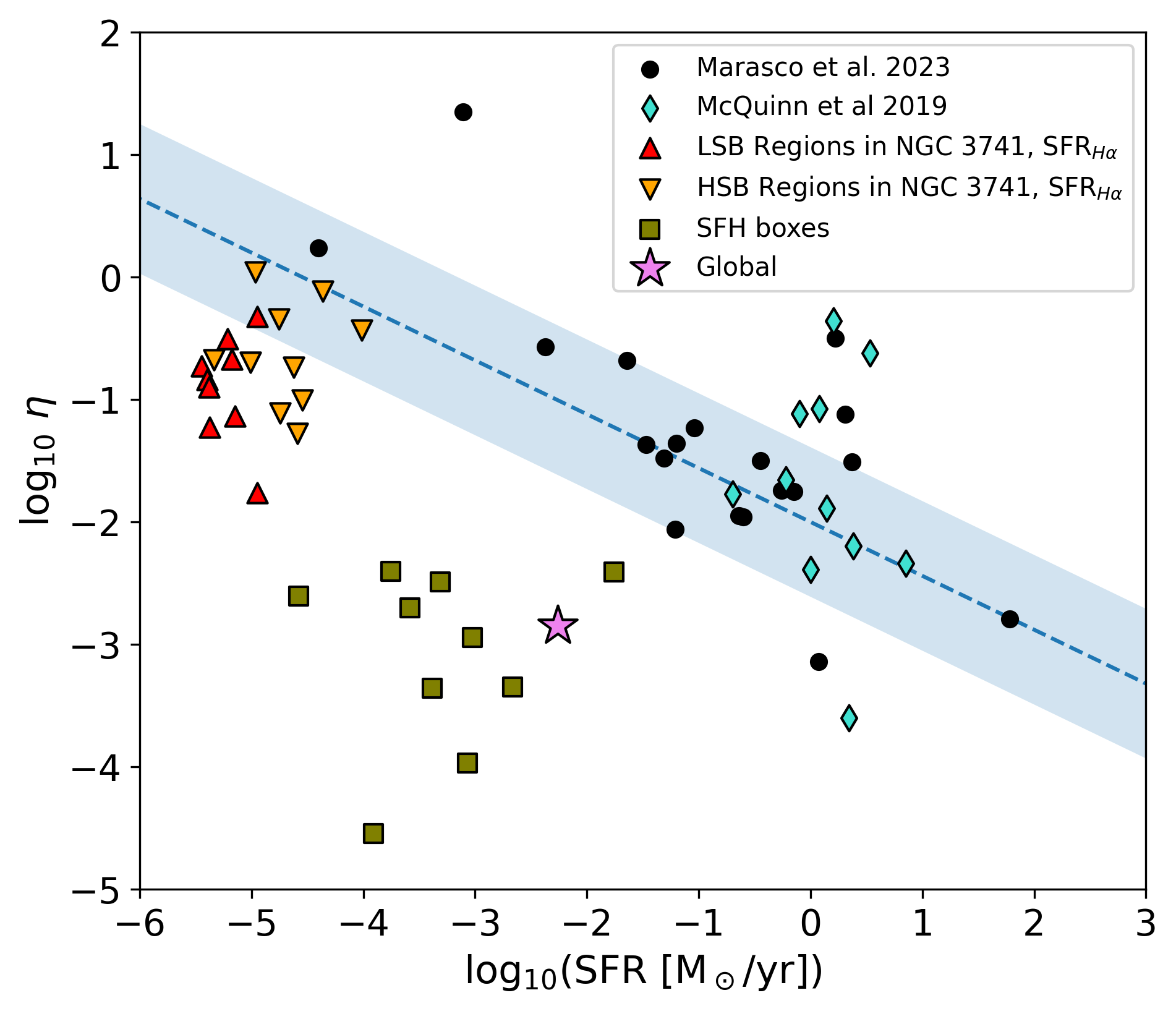}
\caption{ Mass-loading factor as a function of star formation rate in log space. Comparison galaxies from \cite{Marasco2023} are shown in black (circles) and \cite{2019ApJ...886...74M} are shown in cyan (diamonds). The blue dashed line shows the best-fit linear relation from \cite{Marasco2023}, with the blue shaded region showing the intrinsic scatter. LSB H$\alpha$ emission regions are shown in red (triangles), HSB H$\alpha$ emission regions are shown in orange (upside-down triangles), SFH regions are shown in green (squares), and the global value for NGC 3741 is shown in pink (star). The SFR(H$\alpha$) measurements are lower limits, resulting in higher mass-loading factors when compared to those calculated with more reliable star formation tracers.
\label{fig:Massloading}
}
\end{figure}

Following the prescription and SFR measurements described above, we calculate local mass-loading factors for the 19 H$\alpha$ emission regions in NGC 3741. For each region, the H$\alpha$ luminosity is estimated based on the region's total H$\alpha$ flux measured from the stacked spectrum. Assuming symmetrical expansion of the wind, we use the semi-major axis (see Table \ref{tab:hiiregprops}) of the region for $r_{wind}$. We use the average H$\alpha$ $-$ \hi velocity offset as an approximation for the wind speed, $v_{wind}$. The resulting mass-loading factors are plotted as red triangles in Figure \ref{fig:Massloading}. 

Because the SFH boxes include multiple H$\alpha$ emission regions and split regions between boxes, we perform the same method of shifting and stacking spectra as described in Section \ref{sec:SparsePakAnalysis} for the fibers that fall into each SFH box. From these box stacked spectra, we estimate L$^{H\alpha}_{wind}$ and {[SII]}$\lambda_{6716}$/[SII]$\lambda_{6731}$, and v$_{wind}$ is determined by the average magnitude of the velocity differences of the fibers in each box.

We explored multiple methods to determine r$_{wind}$ for each SFH box. For this, we ignore the distinct ionized gas structures and assume that all winds originate from the central concentrated star formation region in the galaxy that lies in SFH box 5. We estimate three radii for each box: 1) the distance from the center of the box to the center of the galaxy, 2) the distance from center of the brightest fiber in the box to the center of the galaxy, and 3) the distance from the center of the fiber with the highest velocity offset in the box to the center of the galaxy. These three radii measurements made only minor changes to the final mass-loading factor calculated for each SFH box, so we report the average value of the three in Table \ref{tab:massload}. The SFH box mass-loading factors are marked with green squares in Figure \ref{fig:Massloading}.

We also estimate a global mass-loading factor for the galaxy. The total H$\alpha$ luminosity of the galaxy was calculated by summing the total flux recovered in all fibers, the {[SII]}$\lambda_{6716}$/[SII]$\lambda_{6731}$ and velocity offset for v$_{wind}$ were both found by taking the average value of all fibers, and the half light radius of the stellar light for r$_{wind}$. We used the global FUV SFR (\citealt{Lee2011}) for the galaxy, and the resulting global mass-loading factor is marked with the pink star in Figure \ref{fig:Massloading}.

The values used to derive the mass-loading factors are given for each region in Table \ref{tab:massload}. The mass-loading factors calculated for each region are plotted as a function of their SFRs against the sample from \cite{Marasco2023} as well as the sample from \cite{2019ApJ...886...74M} in Figure \ref{fig:Massloading}. The mass-loading factors we find for both the H$\alpha$ emission regions and the SFH boxes tend to have lower mass-loading factors than the galaxies in the comparison samples. This indicates that across the galaxy, there is no or negligible mass loss occurring.

\begin{table*}[p]
\caption{Mass-loading Factor Region Values}
\footnotesize
\begin{center}
    \begin{tabular}{l c c c c c c} 
    \hline\hline 
    Region ID &  $L^{H\alpha}_{wind}$  & log(SFR) & {[SII]}$\lambda_{6716}$/[SII]$\lambda_{6731}$ & $n_e$  & v$_{wind}$ &log($\eta$) \\
     &  $\times10^{35}$  (erg s$^{-1}$) & (M$_\odot$ yr$^{-1}$)&  & (cm$^{-3})$ & (km s$^{-1}$) & \\
    \hline 
    1 & 8.36 & -5.18 & 1.29 &113&11.42&-0.67\\
    2 & 14.11 & -4.95& 0.82 &1037&8.28&-1.76\\
    3 & 4.51 & -5.45 &  3.16 &100&4.78&-0.72\\
    4 & 14.12 & -4.95 & 1.37 &100&8.33&-0.32\\
    5 & 5.31 & -5.38 & 1.21 &183&2.06&-1.22\\
    6 & 7.68 & -5.22 & 1.45 &100&6.09&-0.51\\
    7 & 5.06 & -5.40 & 1.27 &135&4.14&-0.84\\
    8 & 5.22 & -5.38 & 1.82 &100&3.28& -0.90\\
    9 & 8.96 & -5.15 & 1.56 &100&2.25&-1.14\\
    10 & 5.78 & -5.34 & 2.94 &100&4.38&-0.67\\
    11 & 12.29 & -5.01 & 1.65 &100&2.80&-0.70\\
    12 & 13.68 & -4.96 &1.39 &100&4.70&0.04\\
    13 & 35.71 & -4.55 & 1.48 &100&1.68&-1.00\\
    14 & 30.01 & -4.62 & 1.43 &100&6.45&-0.73\\
    15 & 22.68 & -4.74 &1.56 &100&2.12&-1.11\\
    16 & 54.92& -4.36 & 1.37 &100&5.40&-0.12\\
    17 & 32.37 & -4.59 & 1.43 &100&0.89&-1.27\\
    18 & 122.55 & -4.01 & 1.38 &100&3.98&-0.43\\
    19 & 22.14 & -4.76 & 1.40 &100&5.86&-0.34\\
    SFH 1 & 4.67 & -4.58 &0.93 &665&14.83&-2.61\\
    SFH 2 & 18.61 & -3.58& 1.02 &455&13.46&-2.70\\
    SFH 3 & 14.02 & -3.76 &  1.91 &100&6.19&-2.40\\
    SFH 4 & 28.30 & -3.02 & 1.69 &100&4.54&-2.94\\
    SFH 5 & 747.99 & -1.76 & 1.44&100&3.24&-2.41\\
    SFH 6 & 37.68 & -2.67 & 1.48 &100&3.88&-3.35\\
    SFH 7 & 4.41 & -3.07 & 1.63 &100&4.39&-3.97\\
    SFH 8 & 2.19 & -3.91 & 0.76 &1297&3.49& -4.54\\
    SFH 9 & 27.52 & -3.31 & 1.28 &129&6.46&-2.49\\
    SFH 10 & 3.23 & -3.38 & 2.24 &100&7.30&-3.36\\
    Global & 888.62 & -2.26 & 1.44 & 100 & 7.78 & -2.85\\
    \hline 
    \end{tabular}
\end{center}

\tablecomments{L$^{H\alpha}_{wind}$ and {[SII]}$\lambda_{6716}$/[SII]$\lambda_{6731}$ were derived using the stacked spectrum from each region. If {[SII]}$\lambda_{6716}$/[SII]$\lambda_{6731}$ $\gtrapprox$ 1.4, the region is at the low density limit of 100 cm$^{-3}$ (\citealt{osterbrock2006}).}
 \label{tab:massload}
\end{table*}

\subsection{Interpreting Mass-loading Factors}
As shown in Figure \ref{fig:Massloading}, the local and the global mass-loading factors calculated for NGC 3741 generally fall below the relation found in for the dwarf galaxies in \cite{Marasco2023}. Both \cite{Marasco2023} and \cite{2019ApJ...886...74M} find that only a small percentage of the gas involved in a wind can potentially escape the virial radius. Most of the material remains within the galactic halo, eventually falling back onto the disk. Given the similarly low or lower mass-loading factors calculated for NGC 3741, the gas is not launched at speed high enough to escape the galaxy. This is confirmed by estimating the escape velocity of the galaxy using $V_{esc} = [2\cdot G \cdot M_{halo}/R_{vir}]^{1/2}$. We estimate an escape velocity of $\sim330$ km s$^{-1}$, where $M_{halo}\sim10^{11} \msun$ (\citealt{Gentile2007}) and $R_{vir} \sim 7.6$ kpc (\citealt{Begum2008}), which is an order of magnitude larger than any wind speed approximations we make for NGC 3741.

While we follow the prescription presented in \cite{Marasco2023} to calculate mass-loading factors, how the winds themselves are actually defined varies. In \cite{Marasco2023}, mass-loading factors are calculated globally for each galaxy, with winds being identified by a broad, low-flux component in the H$\alpha$ emission line and their physical size being the half-light radius of the wind component. Additionally, they calculate FUV and NUV SFRs depending on what observations are available for a given galaxy in their sample. Mass-loading factors in \cite{2019ApJ...886...74M} are calculated globally for each galaxy and are estimated using H$\alpha$ and \hi imaging, with winds and fountains being identified by H$\alpha$ emission extending past the \hi radius of the galaxy. The SFRs for their sample are found based on star formation histories. 

For NGC 3741, we define winds to include the total H$\alpha$ flux within a defined region with radii that correspond to the physical extent of the wind. If we were to adjust these assumptions to match more directly with how winds are defined in \cite{Marasco2023}, the resulting mass-loading factors would be even lower. Additionally, we assume the velocity offsets between the neutral and ionized gas are line-of-sight motions of the ionized gas. However, these offsets could be motions of the \hi gas, which would result in an overestimation of already extremely modest wind speeds in NGC 3741.

When considering the differences between the local mass-loading factors we calculated for the emission regions and SFH boxes, the mass-loading factors for the H$\alpha$ emission regions are higher. This is likely nonphysical, but rather can be attributed to the H$\alpha$ luminosity being an unreliable tracer of star formation activities at such low SFRs. Based on the SFHs, which trace the underlying stellar population, the SFR(H$\alpha$) is an underestimate, resulting in slightly larger mass-loading factors in those regions. Differences in mass-loading factors between the HSB and LSB H$\alpha$ emission regions can be attributed to the smaller radii and generally higher H$\alpha$ luminosities of the HSB regions.

While we have shown that mass-loading factors are sensitive to the method in which the wind is defined, regardless of method, we estimate very low mass-loading factors across NGC 3741, indicative of no or negligible mass loss occurring in the galaxy. Futhermore, these speeds are significantly below the escape velocity of the galaxy, $\sim$330 km s$^{-1}$, and thus, likely do not have enough kinetic energy to carry mass out of the galaxy.

\subsection{Expectations for Outflows From Star Formation Activity}\label{subsec:SFHs}
As stellar feedback drives outflows and consequently the expected mass-loading factors for galaxies, we explore the star formation activity across time for the regions in NGC 3741. Figure \ref{SFHs} shows the local SFHs along with their corresponding spatial locations on the galaxy. In the SFH boxes associated with the diffuse ionized gas features, there is very little star formation activity across the galaxy's lifetime. These regions have, at maximum across time, a peak star formation rate of 3.1 $\times 10^{-4}$ M$_\odot$ yr$^{-1}$. These episodes are fairly short lived and of low amplitude, injecting very little energy into the surrounding ISM across time, aligning with the very low mass-loading rates measured across the galaxy. 
 
However, the central region of the galaxy appears to have experienced a strong starburst in comparison to the rest of the galaxy that began about 40 Myr ago. The SFH box that aligns with the high-surface brightness region peaks at 1.75 $\times 10^{-2}$ M$_\odot$ yr$^{-1}$, which is $\sim$60 times the highest star formation rate in the diffuse regions. Taking into account the total stellar mass in each region, the specific SFR (SFR/M$_*$) in the central region is an order of magnitude larger than that in the diffuse regions.

Given that O stars have typical lifespans of a few million years, there has been enough time for these stars to explode and impact the gas kinematics of the surrounding ISM. Using the SFH, we estimate the total stellar mass formed in this region during the starburst to be 1.7$\times 10^5 \msun$. We can estimate the total number of O stars using this mass and a Kroupa IMF (\citealt{kroupa2001}), which comes out to roughly 500 O stars. For the most significant stellar-feedback-driven outflows observed in low-mass galaxies, the winds are typically driven by energy input from 1000's of SN events (e.g., \citealt{Martin1998}; \citealt{dehorta2014}). Based on the low velocities seen in the diffuse ionized gas and the likely insufficient energy injection from SNe in these regions, we suggest that, though NGC 3741 is in a relative starburst event, the SFR is too low to produce a significant amount of high mass stars to drive outflows and mass loss. Though, we note that outflows are also sensitive to the structure of the surrounding ISM and the timescales of feedback, so there may be more factors behind the lack of significant outflows in NGC 3741.

\section{Conclusions}
\label{sec:conclusion}
In this study, we present SparsePak IFU observations of NGC 3741 and use the spectroscopic data in combination with optical imaging, archival VLA observations, and archival HST observations to understand the impacts of stellar feedback on the ISM. This study serves to establish the methods of analysis of a larger study of the impact of stellar feedback on the ISM in a full sample of 40 nearby low-mass galaxies. NGC 3741 is well matched to the SparsePak field of view and is an intriguing galaxy due to its extremely extended \hi disk. 

From our analysis, we identify 20 regions of H$\alpha$ emission in NGC 3741, and the derived \hii region luminosity function with a power-law index consistent with other studies of low-mass galaxies. The SparsePak spectral analysis revealed that there is a clear difference in [SII]/H$\alpha$ between the low and high-surface brightness regions that could be indicative of different ionization mechanisms in these regions. The spectra for NGC 3741 do not show significant wind components, and the velocity differences between the neutral and ionized gas are moderate. Given these observations, we estimate very low mass-loading factors, which suggests that NGC 3741 has experiences very little or no mass loss due to stellar feedback. This is surprising given the relative starburst event occurring in the center of the galaxy, but we suggest that the SFR is not high enough to produce significant high-mass stars to drive feedback.

As this project continues, this analysis will be performed on the larger, 40 galaxy sample. This sample contains low-mass galaxies with a range of masses and metallicities, which will allow us to explore the impacts of stellar feedback throughout the dwarf galaxy regime. In addition, we can investigate how mass-loading factors may vary on local scales and can connect this to the star formation activity across time for the galaxies in our sample.

\acknowledgments

{\bf Acknowledgments}

This work was conceptualized and made possible by the invaluable contributions of Dr. Liese van Zee who passed February 13, 2024. Dr. van Zee acquired all data taken with the WIYN 3.5m and performed all data reduction and processing for the pODI imaging and SparsePak spectroscopy. Lexi N. Gault acknowledges Liese's mentorship and support as a thesis advisor, and the continuation of this research is done in her honor.  

Lexi N. Gault would like to thank John Salzer for his insightful and helpful feedback on this paper. This work was partially funded by support from the Indiana Space Grant Consortium Fellowship program.

The authors acknowledge the observational and technical support
from the National Radio Astronomy Observatory (NRAO), and
from Kitt Peak National Observatory (KPNO). Observations
reported here were obtained with WIYN 3.5 m telescope which
is a joint partnership of the NSF’s National Optical–Infrared
Astronomy Research Laboratory, Indiana University, the
University of Wisconsin–Madison, Pennsylvania State University, the University of Missouri, the University of
California–Irvine, and Purdue University. This research made
use of the NASA Astrophysics Data System Bibliographic
Services and the NASA/IPAC Extragalactic Database (NED),
which is operated by the Jet Propulsion Laboratory, California
Institute of Technology, under contract with the National
Aeronautics and Space Administration.

\textit{Facility:} the WIYN Observatory; the Very Large Array; Hubble Space Telescope.

\textit{Software:} Astropy (\citealt{astropy,astropy:2018,astropy:2022}); specutils (\citealt{specutils}); astrodendro (\citealt{astrodendro}); lmfit (\citealt{lmfit}); Peak Analysis (\citealt{PANDimeo}); IRAF (\citealt{iraf1986})





\bibliography{mybib}
\bibliographystyle{aasjournal}



\end{document}